\documentclass[11pt]{article}

\usepackage[T1]{fontenc}
\usepackage[utf8]{inputenc}
\usepackage{mathptmx}
\usepackage[a4paper,margin=2.5cm]{geometry}

\usepackage{amsmath}
\usepackage{amssymb}
\usepackage{graphicx}
\usepackage{booktabs}
\usepackage{multirow}
\usepackage{placeins}
\usepackage{authblk}
\usepackage[font=small,labelfont=bf]{caption}
\usepackage[numbers,sort&compress]{natbib}
\usepackage[colorlinks=true,allcolors=blue,breaklinks=true]{hyperref}

\graphicspath{{figures/}}

\title{\bfseries From populations to absolute binding constants in molecular simulations: exact volumetric terms and practical estimators}

\author[1]{Davide Mandelli}
\author[1]{Emiliano Ippoliti}
\author[2]{Charles Plate}
\affil[1]{Institute of Neuroscience and Medicine (INM-9), Computational Biomedicine,
          Forschungszentrum J\"ulich, 52428 J\"ulich, Germany}
\affil[2]{Department of Chemical Engineering, Virginia Tech, Blacksburg,
          Virginia 24061, United States}

\date{}

\begin{document}

\maketitle

\begin{abstract}
Biomolecular simulations of binding commonly confine the unbound ligand with volumetric restraints to accelerate convergence. However, a general account of how such restraints enter the volumetric term of the standard binding free energy, and of how this term should be evaluated from histogram data, is currently lacking. We take as our starting point the established relation, in the dilute limit, between the binding constant $K$ and the relative populations of the bound and unbound states, which we re-derive from the canonical partition function of molecular solutes in explicit solvent. Building on this relation, we show that a restraint that leaves the bound state unaffected and confines a homogeneous unbound region modifies the volumetric term in a single way, by replacing the system volume with the volume of the region over which the unbound population is evaluated. This yields closed-form estimators for restrained simulations, computable from histograms of any suitably defined reaction coordinate, with the volumetric term kept explicit. Previous expressions based on the potential of mean force are recovered as a special case. The analysis also exposes two distinct errors in current practice: single-bin estimators, which reduce the bound state to one histogram bin and are not valid estimators of $K$, and working formulas that omit the bin width entering the volumetric term, which make the result depend on the unit of length. We apply the estimators to a Lennard-Jones dimer, the cucurbit[7]uril/1-adamantanol host--guest complex, and the galactonate--DgoT ligand--protein complex. In both molecular complexes, single-bin estimators underestimate the affinity by $\approx 1$~kcal/mol. Separately, for the structured DgoT bound state, the placement of the state boundary shifts the affinity by $\approx 0.6$~kcal/mol. By anchoring this definition to what experiments resolve, the theory turns it from a hidden assumption into a controlled input, and provides a principled route to absolute binding affinities from molecular simulations.
\end{abstract}



%
\section{Introduction}
Quantifying the thermodynamic stability of non-covalent molecular complexes is a central problem in chemistry, biophysics, and pharmacology~\cite{Gilson2007,Mobley2017}. The equilibrium binding constant $K$ sets the populations of the bound and unbound states at a given concentration and is thus a fundamental physical descriptor of molecular recognition, relevant far beyond drug design, from enzyme regulation to self-assembly. Within pharmacology, binding affinity is among the primary determinants of drug efficacy. Although binding kinetics has emerged as an equally decisive one~\cite{Swinney2004,Copeland2006,Copeland2016,Tonge2018,Bernetti2019}, only affinity and kinetics together can provide the complete description of drug--target engagement~\cite{Decherchi2020}.

The theoretical foundations of binding equilibrium in solution were established decades ago. Seminal contributions~\cite{Chandler1976,Jorgensen1989,Roux1996,Gilson1997,Hermans1997} grounded the problem in the language of statistical mechanics, expressing $K$ in terms of configurational integrals, and established the conceptual framework that continues to underpin modern computational approaches (see Ref.~\cite{Macuglia2025} for a historical account of alchemical methods). Propelled by the rapid growth of high-performance computing, binding free energy calculations have matured to the point of being embedded in drug discovery pipelines~\cite{Rizzi2025}. A rich methodological landscape has accompanied this progress, spanning the alchemical route, in which the ligand is decoupled from its receptor and reinserted into bulk solvent along a nonphysical path via a decoupling parameter $\lambda$ (e.g., free energy perturbation or thermodynamic integration approaches), and the geometrical route, in which the ligand is separated from its partner along a physical path, typically with the help of enhanced-sampling techniques~\cite{Deng2009,Gumbart2013,Mobley2017,Decherchi2020,Fu2022}. These methods are used for both relative and absolute binding free energy calculations. The former estimate affinity differences between ligands~\cite{Cournia2017}, while the latter target the standard binding free energy of individual ligands directly~\cite{Gumbart2013,Fu2022}.

Here we focus on the absolute binding free energy, which yields the binding constant $K$ itself. Whatever the route, a term fixing the standard-state reference is required, whose necessity and origin in the translational entropy of the solutes were spelled out in a classic note by Janin~\cite{Janin1996}. In the alchemical route this term is introduced through the free energy difference between the restrained decoupled ligand and the unrestrained decoupled ligand defined at the standard-state concentration~\cite{Hermans1986,Roux1996,Gilson1997,Hermans1997,DuboueDijon2021}. 
The present work is restricted to the geometrical route, where it takes the form of a volumetric contribution of the unbound state, accounting for the configurational volume accessible to the free ligand. Much of the foundational work on this route is framed in terms of the potential of mean force (PMF)~\cite{Kirkwood1935} along a suitably chosen reaction coordinate~\cite{Jorgensen1989,Roux1999,Allen2004,Woo2005,Trzesniak2007,Deng2009,Gumbart2013,Limongelli2013,Capelli2019,Fu2022}, and these methods have been successfully applied across a wide range of molecular recognition problems~\cite{Decherchi2020}. Self-contained treatments exist for staged protocols, in which restraints are applied and then released through separate free-energy calculations~\cite{Woo2005,Deng2009,Gumbart2013,Fu2022}.

An alternative is to read the affinity off a histogram of a chosen reaction coordinate, sampled in a single simulation in which the ligand reversibly binds and unbinds~\cite{Limongelli2013,Capelli2019,Souza2020,DiMarino2023}, so that the free energy difference follows from the relative populations of the bound and unbound states. The underlying relation between the binding constant and the ratio of these populations is well established~\cite{Roux1996,Woo2005,Deng2009}. For unbiased simulations, De Jong \textit{et al.}~\cite{deJong2011} showed, by thermodynamic arguments and by a statistical-mechanical treatment of dimerizing particles, that for one pair of molecules the binding constant is given by the ratio of the bound and unbound populations multiplied by the accessible volume, and extended the analysis to finite concentrations. Jost Lopez \textit{et al.}~\cite{JostLopez2020} showed that the ratio of populations multiplied by the full box volume carries a finite-size error due to non-specific interactions, which vanishes when only configurations beyond the range of interaction are counted as unbound, together with the corresponding volume. Dubou\'e-Dijon and H\'enin~\cite{DuboueDijon2021} connected this population-based relation to the PMF expression and analyzed the definition of the bound state.

To accelerate convergence, simulations of this kind commonly confine the unbound ligand with volumetric restraints~\cite{Allen2004,Limongelli2013,Capelli2019,DiMarino2023}. In this setting, a general account of how such restraints enter it, and of how it should be evaluated from histogram data, is lacking. Which volume enters the standard-state reference, how the bound and unbound states are delimited on a histogram, and how the bin width enters the result are questions a practitioner must answer, and they are not addressed together in the available, method-specific derivations.

In this work, we address them within a single framework. We take as our starting point the established relation, in the dilute limit, between $K$ and the statistical weights of the bound and unbound states, which we re-derive from the canonical partition function of molecular solutes in explicit solvent. Building on this relation, we show that a restraint that leaves the bound state unaffected and confines a homogeneous unbound region modifies the volumetric term in a single way, by replacing the system volume with the volume of the region over which the unbound population is evaluated. If the restraint also acts on the bound state, a correction term appears, which we give in general form. This yields closed-form practical estimators, computable from histograms of any suitably defined reaction coordinate, in which the unbound state is identified by the relative position of the two molecules while the bound state can be defined according to whatever criterion is appropriate for the system at hand.

The resulting framework encompasses the PMF-based expressions in the literature~\cite{Jorgensen1989,Allen2004,Woo2005,deJong2011,Limongelli2013,DuboueDijon2021} as special cases, which are obtained when the bound and the unbound states are both defined along the same one-dimensional coordinate. At the same time, making every term explicit helps clarifying why a single-bin estimator currently in use~\cite{Sun2014,Comitani2016,Zlobin2021,Szlenk2021,Kuzovlev2023,Borowska2024} is not a valid estimator of $K$, and how to correct working formulas~\cite{Limongelli2013,Souza2020,Lamprakis2021} that are dimensionally inconsistent as printed.

We apply our formalism to three systems of increasing complexity. A Lennard-Jones dimer in explicit solvent serves as the simplest nontrivial testbed. The cucurbit[7]uril--1-adamantanol host--guest complex, whose chemistry underpins biomedical applications~\cite{Aleskovic2024}, and is commonly adopted as a benchmark for binding free energy methods~\cite{Hudson2018,Sun2024,Vijay2025}. The binding of galactonate to the bacterial DgoT transporter, as a real-world problem of current interest~\cite{Liu2021,Dmitrieva2024,Plate2026,Wallis2026}. For each system, we show quantitatively how the choice of the estimator and the definition of the bound state affect the computed binding constant.

\section{Theory}
\subsection{Statistical mechanics of 
chemical equilibrium}
Let us consider the canonical ensemble of a generic system of $N$ particles described by a classical Hamiltonian $\mathcal{H}(q,p)$. Here, $q$ and $p$ indicate the $3N$-dimensional vectors of the canonical coordinates and their conjugate momenta. The thermodynamics of the system can be studied starting from the partition function
\begin{equation}
\label{eq.Z}
    \mathcal{Z}(N,V,T)= \frac{1}{h^{3N}{N}!}
    \int e^{-\beta \mathcal{H}(q,p)}\,{\rm d}q\,{\rm d}p,
\end{equation}
by analyzing the behavior of the Helmholtz free energy\footnote{Free energies are defined up to an additive constant, omitted throughout for brevity.}
\begin{equation}
\label{eq.A}
    F(N,V,T)=-k_{\rm B}T\log\mathcal{Z}
\end{equation}
as a function of the thermodynamic parameters. Here $h$ is Planck's constant, and $\beta=1/k_{\rm B}T$. It is convenient to introduce the canonical probability density function $\rho(q,p)$ to find the system in a
microscopic state $(q,p)$,
\begin{equation}
\label{eq.rhoqp}
    \rho(q,p)=\frac{e^{-\beta \mathcal{H}(q,p)}}
    {\mathcal{N}},\qquad\mathcal{N}=\int e^{-\beta \mathcal{H}(q,p)}\,{\rm d}q\,{\rm d}p.
\end{equation}
$\rho(q,p)$ is normalized,
\begin{equation}
    \int\rho(q,p)\,{\rm d}q\,{\rm d}p=1.
\end{equation}
We also define the statistical weight $P_s$ of a generic
thermodynamic state $s$,
\begin{equation}
\label{eq.Ps}
    P_s=\int\rho(q,p)\,I_s(q,p)\,{\rm d}q\,{\rm d}p,
\end{equation}
as the integral of $\rho$ over the region of phase space that defines that state, expressed here using the indicator function $I_s(q,p)=1$ if $(q,p)\in s$ and zero otherwise. The associated Helmholtz free energy is given by
\begin{equation}
\label{eq.As}
    F_s=-k_{\rm B}T\log P_s,
\end{equation}
and free energy differences between two states by
\begin{equation}
\label{eq.DeltaA}
    \Delta F_{1\to 2}\equiv F_{2}-F_{1}=-k_{\rm B}T\log\frac{P_{2}}{P_1}.
\end{equation}

We now turn to the specific problem of the thermodynamic equilibrium
\begin{equation}
    A+B\rightleftharpoons AB
\end{equation}
of a solution in which two species $A$ and $B$ can form a complex $AB$. In the limit of high dilution, the binding constant
\begin{equation}
    K=\frac{[AB]}{[A][B]}
\end{equation}
defines the thermodynamic equilibrium in terms of the equilibrium concentrations $[\cdot]$. In this context, the thermodynamic states of interest are the bound state ($b$) and the unbound state ($u$), identified by the respective indicator functions $I_{u,b}(q,p)$. The following equation holds:
\begin{equation}
\label{eq.cornerstone}
    K=\frac{P_b}{P_u/V},
\end{equation}
where $P_b$ and $P_u$ are the statistical weights of the bound and unbound states, respectively, as defined in Eq.~\eqref{eq.Ps}, for a system containing one molecule of $A$ and one molecule of $B$ solvated in a volume $V$.
Equation \eqref{eq.cornerstone} is the central relation of this paper. Its
derivation, which starts from the partition function of any number of ligands and receptors  of Eq.~\eqref{eq.Z}, is reported in Section~\ref{subsec:statmec} of the Supporting Information (SI)\footnote{Eq.~\eqref{eq.cornerstone} is equivalent to Eq.~(2) of Ref.~\cite{Woo2005}
(see also Refs.~\cite{Roux1996,Deng2009}), and to the expressions reported in Refs.~\cite{deJong2011,DuboueDijon2021}.}. 

Following an established convention, the binding constant may be reported as a free energy difference
\begin{equation}
\label{eq.dgi0}
    \Delta F_0=-k_{\rm B}T\log\left(KC_0\right),
\end{equation}
at the standard concentration $C_0=1$~mol / litre. This leads to the general
equation~\footnote{Since our derivation is carried out formally in the canonical ensemble, we consistently denote the binding free energy by the Helmholtz free energy $F$. As all results remain unchanged for the case of the isothermal--isobaric ensemble, our $\Delta F_0$ is thus the binding free energy commonly reported as $\Delta G_0$ in terms of the Gibbs free energy $G$.}
\begin{equation}
\label{eq.dgi0corner}
    \Delta F_0=\Delta F_{u\to b}-k_{\rm B}T\log\left(VC_0\right),
\end{equation}
where $\Delta F_{u\to b}=-k_{\rm B}T\log\left(P_b/P_u\right)$ is the volume-dependent free energy difference between the bound and unbound states, and $-k_{\rm B}T\log(VC_0)$ is the volumetric correction. $\Delta F_0$ has a clear physical interpretation (see Section~\ref{subsec:statmec} in the SI): it is the difference between the standard chemical potential of the complex and the sum of those of its partners~\cite{deJong2011}~\footnote{This is equivalent to the reversible work of converting two independently dissolved molecules, each in the standard state, into one standard-state complex, thereby accounting for the loss of translational entropy associated with binding.}.
\subsection{Binding constants from simulations with volumetric restraints}
\label{sec:restraints}
Eq.~\eqref{eq.cornerstone} defines $K$ exactly in terms of the statistical weights of the bound and unbound states. Here, we follow the geometrical route, in which the ratio $P_b/P_u$ is obtained by sampling the reversible transition between the two states along a physical path. Sampling this exchange is the principal bottleneck in practice. Enhanced sampling methods can accelerate this exchange while preserving the ability to recover unbiased statistical weights~\cite{Kumar1992,Torrie1977,Sugita1999,Darve2008,Valsson2016,Lesage2017,Invernizzi2020b,Rizzi2023}. A complementary and widely used strategy is to introduce restraints that limit the configurational volume accessible to the unbound ligand, which would otherwise have to be sampled extensively, so that the ligand returns to the binding site more frequently~\cite{Allen2004,Limongelli2013,Capelli2019}. In this section we derive from Eq.~\eqref{eq.cornerstone} the general expressions for $K$ in the presence of such restraints.

Let us consider an unrestrained system and a restrained one, in which a reflective wall confines the relative position of the two molecules within a region $\Gamma_R$\footnote{In practice, molecular simulations -- including those reported here -- do not employ ideal reflective walls but rather restraining potentials, typically harmonic. For the idealization to hold, the restraining potential must vanish throughout the region of interest and act only within a thin shell at its boundary.}. Alongside the unrestrained density $\rho(q,p)$ of Eq.~\eqref{eq.rhoqp}, we define the restrained canonical probability density
\begin{equation}
\label{eq:rhoR}
    \rho_R(q,p)= \frac{e^{-\beta \mathcal{H}(q,p)}\,I_{R}(q,p)}{\mathcal{N}_R},
    \qquad
    \mathcal{N}_R= \int e^{-\beta \mathcal{H}(q,p)}\, I_{R}(q,p)\,{\rm d}q\,{\rm d}p,
\end{equation}
where $I_R$ is the indicator function of $\Gamma_R$. The two densities coincide up to a multiplication factor inside $\Gamma_R$:
\begin{equation}
\label{eq:rhorel}
    \rho(q,p)=\frac{\mathcal{N}_R}{\mathcal{N}}\,\rho_R(q,p)\qquad \forall\,(q,p)\in\Gamma_R .
\end{equation}

\paragraph*{Statistical weight of the bound state.}
If $\Gamma_R$ contains the entire bound state, Eq.~\eqref{eq:rhorel} holds over it, and integrating over the bound region gives
\begin{equation}
\label{eq:PbR}
    P_b=\frac{\mathcal{N}_R}{\mathcal{N}}\,P_{R,b},
    \qquad
    P_{R,b}=\int \rho_R(q,p)\,I_b(q,p)\,{\rm d}q\,{\rm d}p .
\end{equation}

\paragraph*{Statistical weight of the unbound state.}
Let $\boldsymbol{\xi}=(x,\xi_2,\xi_3)$ be generalized coordinates that specify the position of the center of mass of one molecule relative to that of the other, with $x$ the reaction coordinate (units of length) that identifies the unbound state through the indicator function $I_u(x)$.  Writing the remaining phase-space variables as $\boldsymbol{\eta}$ and the Jacobian determinant of the transformation from $\boldsymbol{\xi}$ to the Cartesian components of the relative position as $J(\boldsymbol{\xi})$, the unrestrained and restrained statistical weights of the unbound state are
\begin{equation}
\label{eq:Pu}
    P_u=\int{\rm d}\boldsymbol{\xi}\,|J(\boldsymbol{\xi})|\,I_u(x)\,\bar\rho(\boldsymbol{\xi}),
    \qquad
    P_{R,u}=\frac{\mathcal{N}}{\mathcal{N}_R}\int{\rm d}\boldsymbol{\xi}\,|J(\boldsymbol{\xi})|\,I_u(x)\,I_R(\boldsymbol{\xi})\,\bar\rho(\boldsymbol{\xi}),
\end{equation}
where $\bar\rho(\boldsymbol{\xi})=\int{\rm d}\boldsymbol{\eta}\,\rho(\boldsymbol{\xi},\boldsymbol{\eta})$ is the density of the relative position per unit Cartesian volume, and Eq.~\eqref{eq:rhorel} has been used for $P_{R,u}$. We now assume that the unbound state is homogeneous, i.e. $\bar\rho(\boldsymbol{\xi})=\bar\rho$, where $\bar\rho$ is a constant\footnote{This condition can be checked in simulations by verifying that the histogram of $x$, divided by the volume of each bin, reaches a plateau in the unbound region.}. Eq.~\eqref{eq:Pu} then gives $P_u=V_u\bar\rho\simeq V\bar\rho$ and $P_{R,u}=(\mathcal{N}/\mathcal{N}_R)\,V_{R,u}\bar\rho$, where
\begin{equation}
\label{eq:VR}
    V_u=\int{\rm d}\boldsymbol{\xi}\,|J(\boldsymbol{\xi})|\,I_u(x),
    \qquad
    V_{R,u}=\int{\rm d}\boldsymbol{\xi}\,|J(\boldsymbol{\xi})|\,I_u(x)\,I_R(\boldsymbol{\xi}),
\end{equation}
are the accessible volume to the unbound state in the unrestrained and restrained cases, respectively\footnote{In the thermodynamic limit $V_u\simeq V$, since the volume of the bound state becomes negligible.}. Hence
\begin{equation}
\label{eq:PuR}
    P_u=\frac{V}{V_{R,u}}\,\frac{\mathcal{N}_R}{\mathcal{N}}\,P_{R,u} .
\end{equation}

\paragraph*{Binding constant.}
Substituting Eqs.~\eqref{eq:PbR} and~\eqref{eq:PuR} into Eq.~\eqref{eq.cornerstone}, the normalization constants cancel and
\begin{equation}
\label{eq:KR}
    K=\frac{P_{R,b}}{P_{R,u}/V_{R,u}},
\end{equation}
or, equivalently,
\begin{equation}
\label{eq:dF0R}
    \Delta F_0 = \Delta F_{R,u \to b} - k_{\rm B}T \log\left(V_{R,u} C_0\right),
    \qquad
    \Delta F_{R,u \to b} = - k_{\rm B}T \log \frac{P_{R,b}}{P_{R,u}}.
\end{equation}
The restraint thus leaves the structure of Eq.~\eqref{eq.dgi0corner} unchanged, with $V$ replaced by $V_{R,u}$. Summarizing, these equations hold under the assumptions of high dilution, homogeneous unbound state and thermodynamic limit (i.e., no finite-size effects~\cite{JostLopez2020}). Since the unbound state extends here over a finite region of the restrained system, we refer to Eq.~\eqref{eq:dF0R} as the extended-region estimator.

Because the unbound state is homogeneous, $P_{R,u}$ can equally be obtained from the density of $x$ at a single point. Let
\begin{equation}
\label{eq:rhof}
    \rho_R(x) = \int \rho_R(q,p)\,\delta\!\left(x - x(q,p)\right)
              \mathrm{d}q\,\mathrm{d}p,
\end{equation}
be the marginal density of $x$ in the restrained system. In the unbound region, Eqs.~\eqref{eq:rhorel} and~\eqref{eq:Pu} give
\begin{equation}
\label{eq:rhoRunb}
    \rho_R(x)=\frac{\mathcal{N}}{\mathcal{N}_R}\,\bar\rho\,S_R(x),
    \qquad
    S_R(x)=\int{\rm d}\xi_2\,{\rm d}\xi_3\,|J(\boldsymbol{\xi})|\,I_R(\boldsymbol{\xi}),
\end{equation}
where $S_R(x)$ is the measure of the section of the restrained region at fixed $x$. Taking the unbound state as the infinitesimal slab $[x_u^*,x_u^*+{\rm d}x)$ within the homogeneous region, $P_{R,u}=\rho_R(x_u^*)\,{\rm d}x$ and $V_{R,u}=S_R(x_u^*)\,{\rm d}x$, and Eq.~\eqref{eq:KR} becomes
\begin{equation}
\label{eq:Khybrid}
    K=S_R(x_u^*)\,\frac{P_{R,b}}{\rho_R(x_u^*)},
\end{equation}
or, equivalently,
\begin{equation}
\label{eq:dF0hybrid}
    \Delta F_0 = -k_{\rm B}T\log\frac{P_{R,b}}{\rho_R(x_u^*)\,{\rm d}x} - k_{\rm B}T\log\left(C_0\,S_R(x_u^*)\,{\rm d}x\right).
\end{equation}
The infinitesimal width ${\rm d}x$ cancels between the two terms, and, since $\rho_R(x)\propto S_R(x)$ in the homogeneous region, the result does not depend on the choice of $x_u^*$ within it. We refer to Eq.~\eqref{eq:Khybrid} as the hybrid estimator, as it combines the statistical weight of the bound state with a pointwise evaluation of the unbound one. The free energy profile enters only through its value in the unbound region. The bound state does not have to be defined in terms of $x$, and $P_{R,b}$ can be evaluated according to whatever definition is appropriate for the system at hand.

Two features of these results deserve emphasis. First, $V_{R,u}$ is the volume of the region over which $P_{R,u}$ is evaluated (that is, the unbound state alone) and not that of the whole restrained region, nor that of the full simulation box. Since $P_{R,u}\propto V_{R,u}$, $K$ does not depend on how much of the homogeneous unbound region is used, down to the infinitesimal slab of the hybrid estimator. The same conclusion has been reached in the context of unrestrained simulations~\cite{deJong2011,JostLopez2020}.

Second, Eq.~\eqref{eq:PbR} assumes that the restraint leaves the bound state untouched. If the restraint also acts on the bound state, Eq.~\eqref{eq:PbR} no longer holds, while the treatment of the unbound state is unchanged. In this case,  Eq.~\eqref{eq:dF0R} generalizes to
\begin{equation}
\label{eq:dF0Rcorr}
    \Delta F_0 = \Delta F_{R,u \to b} + \Delta F_{\rm corr} - k_{\rm B}T \log\left(V_{R,u} C_0\right),
    \qquad
    \Delta F_{\rm corr}=-k_{\rm B}T\log\left(\frac{\mathcal{N}}{\mathcal{N}_R}\,\frac{P_b}{P_{R,b}}\right).
\end{equation}
$\mathcal{N}P_b$ and $\mathcal{N}_RP_{R,b}$ are the configurational integrals over the bound state without and with the restraint, respectively, so that
\begin{equation}
\label{eq:dFcorrav}
    \Delta F_{\rm corr}=-k_{\rm B}T\log\left\langle e^{\beta U_R}\right\rangle_{R,b}\le 0,
\end{equation}
where $U_R$ is the restraining potential and the average is taken over the bound configurations of the restrained simulation. $\Delta F_{\rm corr}$ is the free energy of releasing the restraint in the bound state. It can be obtained through separate free-energy calculations in which the restraint is switched on and off~\cite{Woo2005,Deng2009} or, in principle, by evaluating Eq.~\eqref{eq:dFcorrav} by reweighting~\cite{DuboueDijon2021}. The latter converges only if the bound state is well sampled in the presence of the restraint, and for a hard wall that excludes part of the bound state Eq.~\eqref{eq:dFcorrav} diverges, reflecting that the excluded configurations cannot be recovered.
\subsection{PMF-based expressions as a special case}
\label{sec:pmf}
If the bound state, too, is defined in terms of $x$, through an indicator function $I_b(x)$, then $P_{R,b}=\int{\rm d}x\,I_b(x)\,\rho_R(x)$. It is then convenient to express $\rho_R(x)$ through the free energy of the infinitesimal slab $[x,x+{\rm d}x)$ which, by Eqs.~\eqref{eq.Ps} and~\eqref{eq.As}, is
\begin{equation}
\label{eq:Fslab}
    F_R(x)=-k_{\rm B}T\log\left[\rho_R(x)\,{\rm d}x\right].
\end{equation}
$F_R(x)$ is the free energy profile, or PMF, along $x$. Being the free energy of a thermodynamic state of infinitesimal extent, it is a free energy in the sense of Eq.~\eqref{eq.As}, and differences between its values at two points do not depend on ${\rm d}x$. In terms of $F_R(x)$, the hybrid estimator, including the correction of Eq.~\eqref{eq:dF0Rcorr}, becomes
\begin{equation}
\label{eq:KPMF}
    K = e^{-\beta\Delta F_{\rm corr}}\,S_R(x_u^*)\int{\rm d}x\,I_b(x)\,e^{-\beta\left[F_R(x)-F_R(x_u^*)\right]}.
\end{equation}
Eq.~\eqref{eq:KPMF} recovers the PMF-based expressions of the literature. For an axial coordinate $x=z$ and a cylindrical restraint of radius $R_{\rm cyl}$, $S_R=\pi R_{\rm cyl}^2$ and Eq.~\eqref{eq:KPMF} is the expression of Refs.~\cite{Allen2004,Woo2005,Limongelli2013}, whose factor $e^{\beta\Delta G_{\rm site}}$ corresponds to $e^{-\beta\Delta F_{\rm corr}}$. For the radial distance $x=r$ inside a spherical restraint that leaves the bound state untouched, $S_R(r)=4\pi r^2$, and writing $\rho_R(r)\propto 4\pi r^2 g_{3,R}(r)$, with the radial distribution function $g_{3,R}$ normalized to one in the unbound region, Eq.~\eqref{eq:KPMF} becomes
\begin{equation}
\label{eq:Kg3}
    K=\int{\rm d}r\,I_b(r)\,4\pi r^2\,g_{3,R}(r),
\end{equation}
the integral of the radial distribution function over the bound state~\cite{Jorgensen1989,deJong2011,DuboueDijon2021}.

Two remarks follow. First, these expressions, derived in the literature along different routes, all follow from the single population-based relation of Eq.~\eqref{eq.cornerstone}, which we derive directly from the canonical partition function and which holds independently of the reaction coordinate and of the geometry of the restraint. Second, Eq.~\eqref{eq:KPMF} requires that the bound state be defined in terms of the same one-dimensional coordinate that identifies the unbound state, whereas Eqs.~\eqref{eq:KR} and~\eqref{eq:Khybrid} do not.

A further simplification repeatedly adopted in the literature~\cite{Sun2014,Comitani2016,Zlobin2021,Szlenk2021,Kuzovlev2023,Borowska2024} consists in replacing the statistical weight of the bound state, $P_{R,b}$, by that of a single slab at a point $x_b^*$, typically the minimum of the free energy profile, $e^{-\beta F_R(x_b^*)}=\rho_R(x_b^*)\,{\rm d}x$. With this choice, Eq.~\eqref{eq:dF0hybrid} then becomes
\begin{equation}
\label{eq:singlebin}
    \Delta F_0 = \Delta F_{\rm corr} + F_R(x_b^*)-F_R(x_u^*) - k_{\rm B}T\log\left(S_R(x_u^*)\,{\rm d}x\,C_0\right).
\end{equation}
We refer to Eq.~\eqref{eq:singlebin} as the single-bin estimator, since in practice ${\rm d}x$ is the width of a histogram bin (Section~\ref{sec:practical}). This expression is not a valid estimator of the binding constant. In fact, while the difference $F_R(x_b^*)-F_R(x_u^*)$ does not depend on ${\rm d}x$, the volumetric term does. It follows that $K$ scales linearly with ${\rm d}x$ and vanishes in the continuum limit. In practice, its result therefore depends on an arbitrary choice of bin width, and it misses part of the bound-state weight whenever the bound state is structured. More precisely, comparing Eq.~\eqref{eq:singlebin} with Eq.~\eqref{eq:dF0hybrid}, the single-bin estimator is the hybrid estimator with $P_{R,b}$ replaced by $\rho_R(x_b^*)\,{\rm d}x$, so that the ratio between the single-bin and hybrid estimator binding constants is $\rho_R(x_b^*)\,{\rm d}x/P_{R,b}\le 1$. The single-bin estimator therefore systematically underestimates $K$, yielding values of $\Delta F_0$ that are too positive (see Section~\ref{sec:applications}). It agrees with the hybrid estimator only if a single bin happens to contain the entire bound state.

A second, independent error affects the volumetric term itself when the width ${\rm d}x$ is dropped from Eq.~\eqref{eq:singlebin}. Omitting it leaves the argument of the logarithm with the dimensions of an inverse length, and the result then depends on an arbitrary choice of the unit of length. This is not a formal subtlety. Working formulas that omit ${\rm d}x$ have appeared in the literature, including in widely used methods. In Ref.~\cite{Limongelli2013}, the integral expression is correct and coincides with Eq.~\eqref{eq:KPMF}, but their working formula, Eq.~(3), writes the volumetric correction as $-k_{\rm B}T\log(\pi R_{\rm cyl}^2C_0)$, which is dimensionally incorrect, missing a length factor. Related inconsistencies are found in expressions reported for the radially symmetric case~\cite{Souza2020,Lamprakis2021}.
\section{Practical estimators for molecular simulations}
\label{sec:practical}
Molecular simulations~\cite{frenkel2023} generate discrete phase-space trajectories. Considering $M$ configurations extracted from a trajectory -- providing a finite number of samples $\{x_i\}_{i=1,\dots,M}$ -- the natural estimator for $\rho(x)$ is the normalized histogram over a discrete mesh $\{x_k\}$ with bins of user-defined width $\Delta x$. Denoting by $M_k$ the number of samples that fall into the $k$th bin, the normalized bin count
\begin{equation}
\label{eq:Pk}
    P_k = \frac{M_k}{M}
        \approx \int_{x_k}^{x_k+\Delta x} \rho(x)\,\mathrm{d}x
\end{equation}
approximates the exact statistical weight of that bin and is dimensionless. Consequently, the statistical weight of a thermodynamic state $s$ is approximated by
\begin{equation}
\label{eq:Ps_disc}
    P_s \approx \sum_{k\in s} P_k,
\end{equation}
where the sum runs over the bins that define the state $s$, consistently with Eq.~\eqref{eq.Ps}. In the ergodic limit and with infinitely fine resolution, the histogram recovers the true density, $\rho(x) = \lim_{M\to+\infty}\lim_{\Delta x\to 0} P_k/\Delta x$. In practice, one always works at finite $M$ and finite $\Delta x$. The free energy profile is estimated as
\begin{equation}
\label{eq:Afmd}
    F(x_k) = -k_{\rm B}T\log P_k,
\end{equation}
the discrete counterpart of Eq.~\eqref{eq:Fslab}, and free energy differences between two states as
\begin{equation}
\label{eq:DA12}
    \Delta F_{1\to 2} =
    -k_{\rm B}T\log\!\left(\frac{\sum_{k\in 2} P_k}{\sum_{k\in 1} P_k}\right)
    = -k_{\rm B}T\log\!\left(
        \frac{\displaystyle\sum_{k\in 2} e^{-\beta F(x_k)}}
             {\displaystyle\sum_{k\in 1} e^{-\beta F(x_k)}}
    \right).
\end{equation}
For restrained simulations, the same expressions apply to the histogram $P_{R,k}$ and to the profile $F_R(x_k)=-k_{\rm B}T\log P_{R,k}$. If the bound state is defined by other coordinates, $P_{R,b}$ is obtained in the same way from the corresponding histogram.

\paragraph*{Estimators.}
With these ingredients, the three estimators of Sections~\ref{sec:restraints} and~\ref{sec:pmf} take the following forms. In each of them, $\Delta F_{\rm corr}$ of Eq.~\eqref{eq:dF0Rcorr} vanishes when the restraint leaves the bound state unaffected.

\textit{Extended-region estimator}, from Eqs.~\eqref{eq:dF0R} and~\eqref{eq:dF0Rcorr}. With the unbound state defined as the interval $x\in[x_u^*,x_u^*+\ell)$ within the homogeneous region,
\begin{equation}
\label{eq:extended}
    \Delta F_0 = \Delta F_{\rm corr} - k_{\rm B}T\log\frac{P_{R,b}}{\sum_{k\in u}P_{R,k}} - k_{\rm B}T \log\left(V_{R,u} C_0\right),
    \qquad
    V_{R,u}=\int_{x_u^*}^{x_u^*+\ell}{\rm d}x\,S_R(x),
\end{equation}
where $V_{R,u}$ follows from Eq.~\eqref{eq:VR}, $S_R(x)$ is defined in Eq.~\eqref{eq:rhoRunb}, and $P_{R,b}=\sum_{k\in b}P_{R,k}$ when the bound state is defined in terms of $x$.

\textit{Hybrid estimator}, from Eq.~\eqref{eq:dF0hybrid}. With the unbound state reduced to the single bin $k^*$ centered at $x_u^*$ of volume $V_{R,u}=S_R(x_u^*)\,\Delta x$,
\begin{equation}
\label{eq:hybridgen}
    \Delta F_0 = \Delta F_{\rm corr} -k_{\rm B}T\log\frac{P_{R,b}}{P_{R,k^*}} - k_{\rm B}T\log\left(S_R(x_u^*)\,\Delta x\,C_0\right).
\end{equation}

\textit{Single-bin estimator}, from Eq.~\eqref{eq:singlebin}. With also the bound state reduced to the single bin centered at $x_b^*$,
\begin{equation}
\label{eq:pointsgen}
    \Delta F_0 = \Delta F_{\rm corr} + F_R(x_{b}^*)-F_R(x_{u}^*)
               - k_{\rm B}T \log\left(S_R(x_u^*)\,\Delta x \, C_0\right).
\end{equation}

\begin{figure}[!t]
    \centering
    \includegraphics[width=0.8\textwidth]{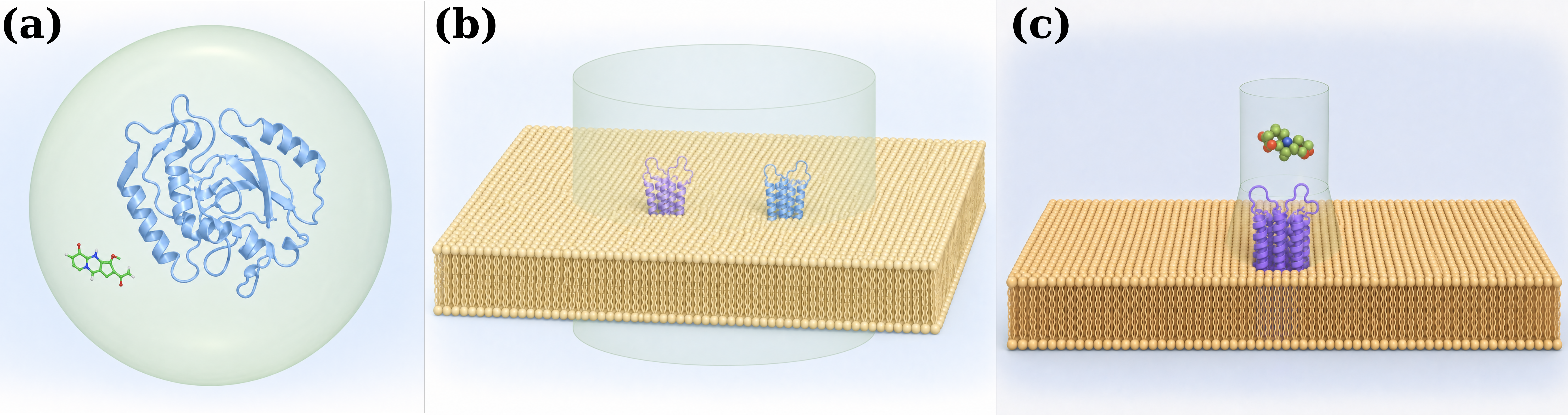}
    \caption{
        (a)~A spherically symmetric case: a small-molecule ligand binds to a soluble protein in aqueous solution in the absence of a well-defined binding pocket.
        (b)~A circularly symmetric case: two membrane-embedded proteins dimerize within a lipid bilayer.
        (c)~A cylindrically symmetric case: a ligand binds to the buried pocket of a membrane-embedded protein from the aqueous phase. In all panels, restraints are shown as translucid green surfaces.
    }
    \label{fig:geometries}
\end{figure}

The bin width $\Delta x$ appears explicitly in the volumetric terms of Eqs.~\eqref{eq:hybridgen} and~\eqref{eq:pointsgen}, as the finite counterpart of the infinitesimal interval ${\rm d}x$ of Eqs.~\eqref{eq:dF0hybrid} and~\eqref{eq:singlebin}. It is not an artefact of the discretization and must be included for the expressions to be dimensionally consistent and numerically correct. In the hybrid estimator it is compensated by $P_{R,k^*}\propto\Delta x$, so that, for sufficiently fine bins, the result does not depend on the bin width. In the single-bin estimator, the difference $F_R(x_b^*)-F_R(x_u^*)$ is itself independent of $\Delta x$, since both bin weights scale with it, and the explicit $\Delta x$ in the volumetric term is left uncompensated. For example, assuming full statistical convergence, halving the bin width raises $\Delta F_0$ by $k_{\rm B}T\log 2$, showing once more that this is a not well-defined estimator.
\paragraph*{Specialization to common geometries.}
While the choice of reaction coordinates that best characterize the bound state is system-dependent, the unbound state can always be identified in terms of the distance between the two reactants, whose appropriate definition is guided by the approximate symmetry of the system. Three geometries are of broad practical relevance in biomolecular simulations (Fig.~\ref{fig:geometries}): the spherically symmetric case, describing association in solution, including ligands interacting with targets that lack well-defined binding pockets; its two-dimensional analog, describing molecules dimerizing within a membrane; and the cylindrical case, describing ligands binding to spatially localized pockets, whether in membrane-embedded proteins, buried active sites of soluble proteins, or supramolecular host--guest systems. The expressions in this Section specialize readily to each of them, through the choice of the reaction coordinate and of the shape of the restraint, which fix $S_R(x)$ and $V_{R,u}$. The exact expressions used to analyse the systems studied in Section~\ref{sec:applications} are given in Section~\ref{sec:compdetails}. Those for all three geometries and all three estimators, including their formulation in terms of radial distribution functions, are derived in Section~\ref{sec:SI_estimators} of the SI.
\section{Computational Details}
\label{sec:compdetails}
Molecular dynamics simulations for the Lennard-Jones (LJ) and the host-guest system were performed de novo for this study. Simulation data for the GAL--DgoT ligand--protein system were taken from Ref.~\cite{Plate2026} . Full simulation settings are reported in Sections~\ref{subsec:LJ}--\ref{subsec:dgot} of SI. In this section we describe how the binding affinities were calculated, using the estimators of Section~\ref{sec:practical} specialized to the geometry of each system.

\paragraph*{Estimating statistical weights from finite sampling}
Practical evaluation of Eq.~\eqref{eq.Ps} from finite simulation trajectories can be challenging in practice. For complex molecular systems, thorough sampling is rarely achieved, and the resulting histograms are often noisy or sparsely populated, which can lead to numerical instabilities. An effective remedy is to replace the raw bin counts with a kernel density estimate~\cite{silverman1986density}, smoothing the histogram over a length scale comparable to the bin width. Using Gaussian kernels of width $\sigma$, the statistical weight of state $s$ can be estimated as
\begin{equation}
    P_s = \frac{1}{\mathcal{N}}
    \sum_{n=1}^{M} w_n \sum_{k \in s}
    \exp\!\left(-\frac{(x_n - x_k)^2}{2\sigma^2}\right),
\end{equation}
where $x_n$ is the value of the reaction coordinate at the $n$th frame, $x_k$ are the bin centers, $w_n$ is the statistical weight of the $n$th frame, and the outer sum runs over all $M$ frames of the trajectory. The normalization constant is 
\begin{equation}
    \mathcal{N} =   \sum_{n=1}^{M} w_n \sum_{k}
    \exp\!\left(-\frac{(x_n - x_k)^2}{2\sigma^2}\right),
\end{equation}
where the index $k$ runs over all bins.\\
For unbiased simulations $w_n = 1$ for all $n$. In the metadynamics simulations used in this work, $w_n = e^{+\beta V_n}$, where $V_n$ is the external bias potential accumulated at the $n$th frame~\cite{Tiwary2015a,Invernizzi2020,Giberti2020}. A common choice for the width of the Gaussian kernel is $\sigma = \Delta x / 2$, where $\Delta x$ is the user-defined bin width of the mesh.
\paragraph*{Lennard-Jones dimer.}
We consider a model system consisting of two particles (P) -- hereafter the dimer -- immersed in a bath of solvent particles (S). All interactions follow the standard 12-6 form
\begin{equation}
    V_{\rm \alpha\beta}(r)=4\varepsilon_{\rm \alpha\beta}\left[\left(\frac{\sigma}{r}\right)^{12}-\left(\frac{\sigma}{r}\right)^6\right]
\end{equation}
with $\varepsilon_{\rm SS} = \varepsilon_{\rm SP} = 1$, $\varepsilon_{\rm PP} = 5$, and $\sigma=1$. The system is simulated at the number density $\rho_0 = 0.8\,\sigma^{-3}$ and temperature $T = 0.9\,\varepsilon_{\rm SS}/k_{\rm B}$, which corresponds to a dense liquid phase of the solvent. The dimer is thermodynamically stable against dissociation at the given thermodynamic conditions.

\begin{figure}[!t]
    \centering
    \includegraphics[width=0.8\textwidth]{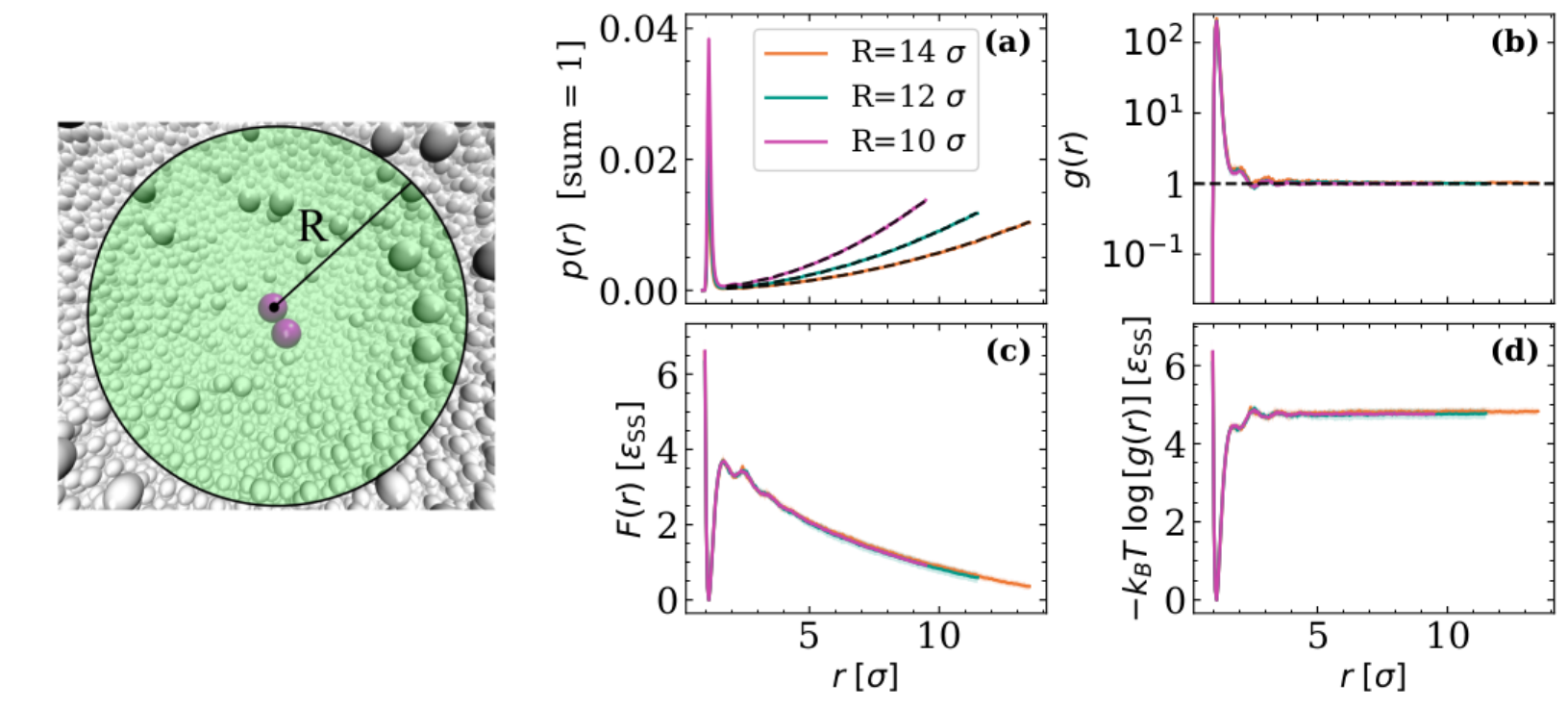}
    \caption{
        (\textit{Left}) View of the system used in simulations. The dimer (magenta) interparticle distance $r$ is restrained within a sphere of radius $R$ (green).
        (\textit{Right}) (a)~Probability histograms $p(r)$. Dashed lines are quadratic fit of $p(r)\propto r^2$ in the region $r>1.8\, \sigma$. (b)~Radial distribution function $g(r)$ on a logarithmic scale. The dashed horizontal line marks the asymptotic value of $g = 1$. (c)~Free energy profile $F(r) = -k_{\rm B}T\log p(r)$, zeroed at its minimum. (d)~Potential of mean force $-k_{\rm B}T\log g(r)$, zeroed at its minimum. In all panels, three curves are shown for the three radii of the spherical restraint $R=10\,\sigma,\,12\,\sigma,\,14\,\sigma$. Shaded bands indicate uncertainties as $\pm 1$ SD from seven independent estimates.}
    \label{fig:LJ}
\end{figure}

Molecular dynamics simulations were performed with one P particle held fixed at the center of the box and the second confined by a spherical harmonic restraint of radius $R$ (Fig.~\ref{fig:LJ}, left panel). Holding one particle fixed does not affect the equilibrium distribution of the inter-particle distance, owing to the translational invariance of the system in the homogeneous bulk. Three values of the restraint radius were considered: $R=10\,\sigma,\, 12\,\sigma,\, 14\,\sigma$. This system maps exactly onto the spherically symmetric case, with the dimer inter-particle distance $r$ as the natural reaction coordinate.

With $x=r$, the restraint contains the bound state and acts only within a thin shell at $R$, so that $\Delta F_{\rm corr}=0$, and the section of the restrained region at fixed $r$ is the spherical surface $S_R(r)=4\pi r^2$. We used the extended-region estimator, Eq.~\eqref{eq:extended}. For an unbound state defined as the shell $r\in[r_u^*,r_u^*+\ell)$ inside the restraining sphere, it reads
\begin{equation}
\label{eq:VRsph}
    \Delta F_0=-k_{\rm B}T\log\frac{\sum_{k\in b}P_{R,k}}{\sum_{k\in u}P_{R,k}}-k_{\rm B}T\log\left(V_{R,u} C_0\right),
    \qquad
    V_{R,u}=\frac{4}{3}\pi\left[(r_u^*+\ell)^3-{r_u^*}^3\right],
\end{equation}
where $V_{R,u}$ is the volume of the unbound shell alone, not that of the whole restraining sphere. Convergence of the unbound region was monitored through the radial distribution function
\begin{equation}
\label{eq.g3}
    g(r_k) =
    \frac{\text{\# samples } r \in [r_k,\,r_k+\Delta r)}{4\pi r_k^2 C_v M \Delta r},
\end{equation}
where $M$ is the total number of samples and $C_v$ is the volumetric concentration of the free particle, chosen such that $g(r) \to 1$ in the homogeneous unbound region (when convergence is reached). An equivalent formulation of Eq.~\eqref{eq:VRsph} in terms of $g$, the hybrid and single-bin estimators for this geometry, and the analogous expressions for the circularly symmetric case are given in Section~\ref{sec:SI_estimators} of the SI.

Probability histograms $p(r)$ and radial distribution functions $g(r)$ were built directly from the sampled configurations by binning, without kernel density estimation. The $g(r)$ was normalized using the nominal concentration  $C_v=1/\widetilde V$ of the free P particle, where $\widetilde V$ is the volume of a sphere corresponding to the maximum value of $r$ sampled during the simulation. Histograms were computed on a uniform grid spanning $r\in[0.85\,\sigma,13.5\,\sigma]$ using a bin width $\Delta r=0.05\,\sigma$. The binding free energy difference was computed as $\Delta F_{R,u\to b} = -k_{\rm B}T \log(N_{R,b} / N_{R,u})$, using directly the numbers $N_{R,b}$ and $N_{R,u}$ of trajectory frames falling in the bound ($r\in[0,\, 2.5\,\sigma$)) and unbound ($r\in[2.5\,\sigma,\, r_{\rm max}$]) regions, respectively, where $r_{\rm max}=R-0.5\,\sigma$. The accessible unbound volume entering the volumetric correction $-k_{\rm B}T \log(V_{R,u} C_0)$ was computed as $V_{R,u} = \frac{4}{3}\pi\left(r_{\rm max}^3-(2.5\,\sigma)^3\right)$, yielding $V_{R,u} \approx 3{,}526\,\sigma^3$, $6{,}305\,\sigma^3$, and $10{,}241\,\sigma^3$, respectively for $R = 10\,\sigma,\, 12\,\sigma$, $14\,\sigma$. The standard binding affinity was computed as $\Delta F_{0} = \Delta F_{R,u \to b} -k_{\rm B}T \log(V_{R,u} C_0)$, with $C_0$ = $\sigma^{-3}$. Statistical uncertainties were estimated by dividing each trajectory into seven contiguous segments of equal length and computing standard deviations across them.
\paragraph*{Cucurbit[7]uril/1-adamantanol host--guest complex.}
Cucurbit[7]uril (CB7) is a barrel-shaped synthetic receptor with a hydrophobic cavity capable of encapsulating a wide range of guest molecules with remarkably high affinity~\cite{Moghaddam2011,Grimm2022}. The binding of 1-adamantanol to the macrocyclic CB7 host in explicit water represents a well-studied benchmark for computational binding free energy methods~\cite{Vijay2025}. 

We performed well-tempered metadynamics~\cite{Barducci2008} simulations of the CB7/1-adamantanol complex in explicit water, confining the center-of-mass of 1-adamantanol within a cylinder of radius $R$ in the unbound region (Fig.~\ref{fig:cb7_results}, left panel). An additional repulsive wall was applied on one side of CB7 to restrict binding to a single portal of the receptor, accelerating convergence without affecting the computed binding affinity. The reaction coordinate used to estimate the binding affinity is the projection $z$ onto the CB7 symmetry axis of the host-guest center-of-mass displacement. To demonstrate the independence of the prediction on the restraint geometry, we tested three different cylinder radii of $R = 0.1,\,0.2,\,0.3~\mathrm{nm}$.

This system is a realization of the cylindrical case, with $x=z$. The restraint confines the lateral displacement of the ligand within a cross-section of area $\Sigma=\pi R^2$, so that $S_R(z)=\Sigma$ in the unbound region, and, since it does not act on the bound state, $\Delta F_{\rm corr}=0$. The probability density $\rho_R(z)$ then  provides the convergence diagnostic itself, since it approaches a plateau far from the binding site. We used all three estimators of Section~\ref{sec:practical}, which specialize as follows. For the extended-region estimator, Eq.~\eqref{eq:extended}, with the unbound state defined as the interval $z\in[z_u^*,z_u^*+\ell)$, $V_{R,u}=\Sigma \ell$ and
\begin{equation}
\label{eq.K_1D_rest}
    \Delta F_0 = -k_{\rm B}T\log\frac{\sum_{k\in b}P_{R,k}}{\sum_{k\in u}P_{R,k}} - k_{\rm B}T \log(\Sigma \ell C_0).
\end{equation}
For the hybrid estimator, Eq.~\eqref{eq:hybridgen}, with the unbound state reduced to the bin of width $\Delta z$ centered at $z_u^*$,
\begin{equation}
\label{eq:hybrid}
    \Delta F_0 = -k_{\rm B}T\log\frac{\sum_{k\in b}P_{R,k}}{P_{R,k^*}}
               - k_{\rm B}T \log\!\left(\Sigma \Delta z \, C_0\right),
\end{equation}
which is the discrete form of the PMF-based expression of Refs.~\cite{Allen2004,Woo2005,Limongelli2013}. For the single-bin estimator, Eq.~\eqref{eq:pointsgen}, with the bound state reduced to the bin centered at $z_b^*$,
\begin{equation}
\label{eq:points}
    \Delta F_0 = \left( F_R(z_{b}^*)-F_R(z_{u}^*)\right)
               - k_{\rm B}T \log\!\left(\Sigma \Delta z \, C_0\right).
\end{equation}
In Table~\ref{tab:cb7_results}, $\Delta F_{R,u\to b}$ denotes the first term on the right-hand side of each of these expressions, and $-k_{\rm B}T\log(V_{R,u}C_0)$ the second.

Probability histograms $p(z)$ were constructed using kernel density estimation on a uniform grid spanning $z \in [-0.3,2.0]$~nm with $N = 100$ bins, corresponding to a bin width $\Delta z = 0.023$~nm, using Gaussian kernels of width $\sigma = \Delta z/2$. Unbiased probability distributions were obtained using the reweighting technique of Ref.~\cite{Tiwary2015a} . The first 100~ns of trajectory, required to reach the adiabatic regime of well-tempered metadynamics, were removed from analysis (see section~\ref{subsec:cb7} in SI). The volume-dependent free energy difference was computed as $\Delta F_{R,u\to b} = -k_{\rm B}T \log(P_{R,b} / P_{R,u})$, where $P_{R,b}$ and $P_{R,u}$ are the statistical weights of the bound ($z \in [-0.1, 0.1]$~nm) and unbound ($z \in [1.5, 1.7]$~nm) regions, respectively. For the hybrid estimator, the unbound region was defined as the bin centered at $z_u^*=1.57$~nm. The volume correction $-k_{\rm B}T \log(V_{R,u} C_0)$ was applied, where $V_{R,u} = \pi R^2 \ell$ is the effective unbound volume, with $R$ the cylinder radius and $\ell$ the extension of the unbound region: $\ell=0.2$~nm for the extended-region estimator, and $\ell=\Delta z$ for both the hybrid- and single-bin estimators. $C_0$ = 1 mol/l $= 0.6022$~nm$^{-3}$ is the standard concentration. For the single-bin estimator, the bound state was defined as the absolute minimum of the free energy, which was always set to zero. The standard binding affinity was computed as $\Delta F_{0} = \Delta F_{R,u \to b} -k_{\rm B}T \log(V_{R,u} C_0)$. Statistical uncertainties were estimated by dividing each trajectory into six contiguous segments of equal length and computing standard deviations across them.
\paragraph*{Galactonate--DgoT ligand--protein complex.}
The bacterial D-galactonate transporter DgoT is a member of the SLC17 family of organic anion transporters, which plays a central role in the loading of neurotransmitters into synaptic vesicles and the export of lysosomal sugars~\cite{Liu2021,Dmitrieva2024,Plate2026,Wallis2026}. Here, we apply our scheme to compute the binding affinity of the substrate galactonate (GAL), in its deprotonated form, dissociating from the inward-facing, gate-open conformation of DgoT, as reported in Ref.~\cite{Plate2026}. The left panel of Fig.~\ref{fig:dgot_results} shows a schematics of the simulation setting, in which well-tempered metadynamics~\cite{Barducci2008} was used in combination with a smooth funnel-shaped confining potential, in the spirit of funnel metadynamics~\cite{Limongelli2013}, to guide GAL along the unbinding $z$-axis of our chosen frame of reference. The reaction coordinate used to estimate the binding affinity is the projection $z$ onto this axis of the center-of-mass displacement of the ligand from the protein binding pocket. Since the confining potential ends in a cylinder of radius $R$ in the unbound region and does not act on the bound state, the cylindrical expressions used for CB7, Eqs.~\eqref{eq.K_1D_rest}--\eqref{eq:points}, apply here as well, with $\Sigma=\pi R^2$.

The probability histogram $p(z)$ was constructed using kernel density estimation on a uniform grid spanning $z \in [-0.3,4.4]$~nm with $N = 150$ bins, corresponding to a bin width $\Delta z = 0.031$~nm, using Gaussian kernels of width $\sigma = \Delta z / 2$. Unbiased probability histograms were obtained using the reweighting technique of Ref.~\cite{Tiwary2015a}. The first $1$~$\mu$s of the trajectory, required to reach the adiabatic regime of well-tempered metadynamics, were removed from analysis. The volume-dependent free energy difference was computed as $\Delta F_{R,u\to b} = -k_{\rm B}T \log(P_{R,b} / P_{R,u})$, where $P_{R,b}$ and $P_{R,u}$ are the statistical weights of the bound ($z \in [-0.2, 0.2]$~nm, for the first two entries of Table~\ref{tab:dgot_results}, or $z \in [-0.2, 3.8)$~nm, for the last entry of Table~\ref{tab:dgot_results}) and unbound ($z \in [3.8, 4.2]$~nm) regions, respectively. For the hybrid estimator the unbound region was defined as the bin at $z_u^*=4.0$~nm. The volume correction $-k_{\rm B}T \log(V_{R,u} C_0)$ was applied, where $V_{R,u} = \pi R^2 \ell$ is the effective unbound volume. $R=0.1$~nm is the radius of the final cylindrical part of the smooth restraint and $\ell$ is the extension of the unbound region: $\ell=0.4$~nm for the extended-region estimator, and $\ell=\Delta z$ for both the hybrid- and single-bin estimators. $C_0 = 1$ mol/l $= 0.6022$~nm$^{-3}$ is the standard concentration. For the single-bin estimator, the bound state was defined as the absolute minimum of the free energy, which was always set to zero. The standard binding affinity was computed as $\Delta F_{0} = \Delta F_{R,u \to b} -k_{\rm B}T \log(V_{R,u} C_0)$. As a qualitative indicator of convergence, we monitored $\Delta F_{R,u\to b}(t)$ as a function of cumulative simulation time $t$ and reported the maximum deviation of this quantity from its best estimate (computed using the full dataset) over the last $4$~$\mu$s of the trajectory (see section~\ref{subsec:dgot} in SI).
\section{Applications}
\label{sec:applications}
\paragraph*{Lennard-Jones dimer.}
The left panel of Fig.~\ref{fig:LJ} shows a schematics of the simulation setting, with the dimer (magenta) immersed in a bath of solvent particles (white) and subjected to a spherical restraining potential of radius $R$ (green). The right panel summarizes the results of the analysis of the trajectories obtained at different values of $R$. Panel (a) shows the probability histograms $p(r)$, which display a sharp peak at $r\approx1.1\,\sigma$, followed by the expected quadratic growth (dashed line). The corresponding free energy profiles (panel (c)) show a well-defined bound minimum followed by a monotonically decreasing region in the unbound state, reflecting the quadratic growth of $p(r)$. Panel (b) displays the radial distribution function $g(r)$ on a logarithmic scale, approaching the asymptotic value of $g(r)\approx 1$ at large distances, demonstrating that the histograms are well converged. Panel (d) shows the corresponding PMF, $-k_{\rm B}T\log g(r)$, which is consistent across the three restraining radii and reaches a flat plateau at large values of $r\gtrsim 4\,\sigma$.

\begin{figure}[!t]
    \centering
    \includegraphics[width=0.7\textwidth]{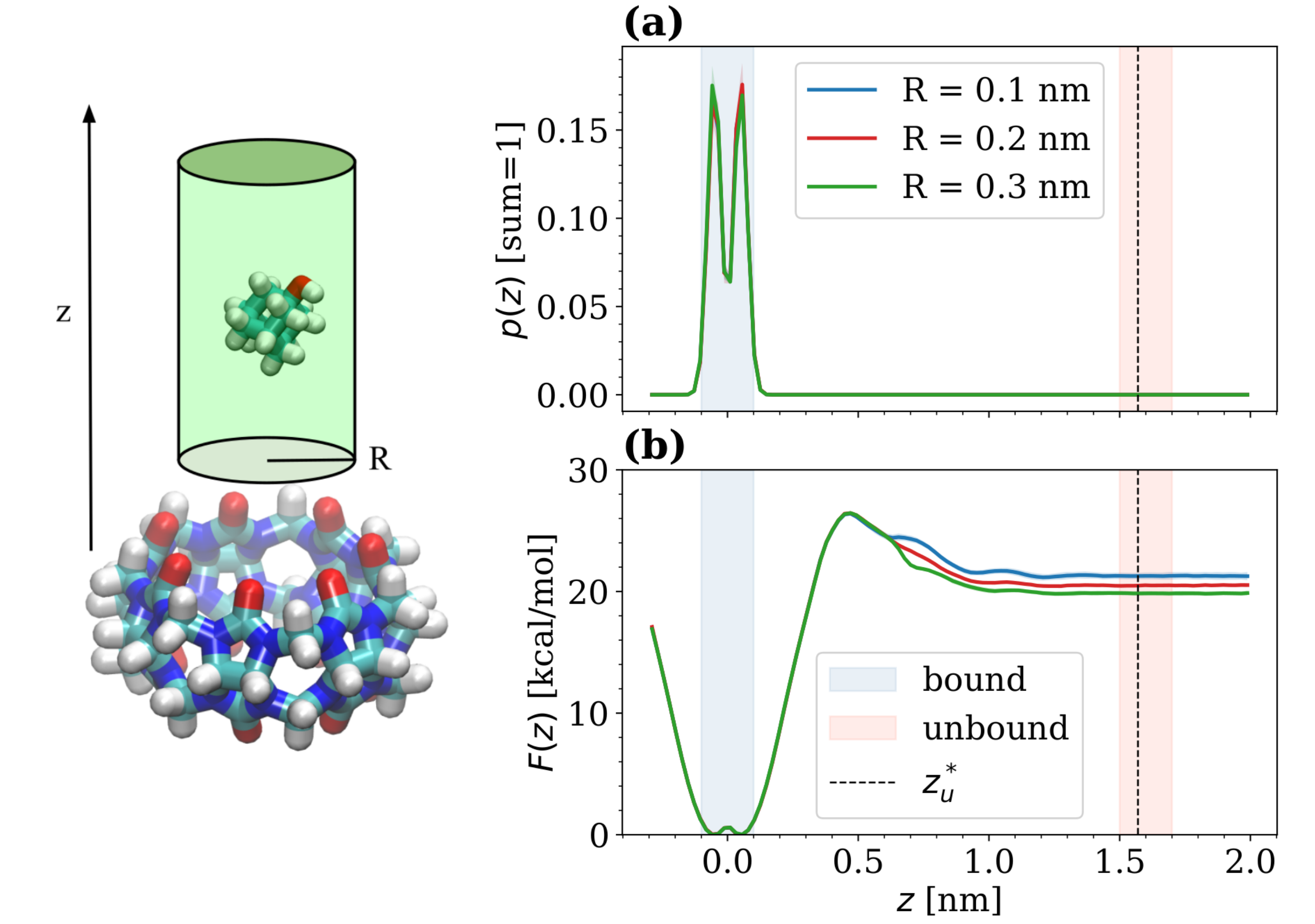}
    \caption{
        (\textit{Left}) Atomistic model of 1-adamantanol (top) and cucurbit[7]uril (bottom) in explicit water (omitted for clarity) used in simulations. Atoms are colored by element: carbon (cyan), oxygen (red), nitrogen (blue), and hydrogen (white). In the unbound state, the center-of-mass of 1-adamantanol is confined to remain within a cylinder of radius $R$ (green). The arrow indicates the $z$ reaction coordinate. (\textit{Right}) (a)~Probability histograms $p(z)$ for three cylinder radii $R=0.1,\, 0.2,\, 0.3~\mathrm{nm}$. (b)~Corresponding free energy profiles $F(z) = -k_{\rm B}T\log p(z)$, zeroed at their respective minima.
        The blue and red shaded regions indicate the bound and unbound integration windows used for the extended-region estimator of $\Delta F_0$. The dashed vertical line marks the single-bin unbound reference $z_u^* = 1.57$~nm. The single-bin bound reference was taken at the absolute minimum of the free energy profile, always set to zero. In all panels, shaded bands indicate $\pm 1$ SD from six independent estimates (bands narrower than the line width almost everywhere).
    }
    \label{fig:cb7_results}
\end{figure}

Table~\ref{tab:LJ} collects the numerical values of the binding affinity $\Delta F_0$. While $\Delta F_{R,u\to b}$ changes by $\approx 0.9\,\varepsilon_{\rm SS}$ across the three restraint radii, the values of $\Delta F_0$ agree within statistical uncertainty, with a maximum spread of $\approx 1\,\%$. This confirms that the volumetric correction $-k_{\rm B}T\log(V_{R,u} C_0)$ accounts for the dependence of $\Delta F_{R,u\to b}$ on the restrained volume.
\begin{table}[!t]
    \centering
    \caption{
        \textbf{Binding affinity of the Lennard-Jones dimer.}
        Energies in units of $\varepsilon_{\rm SS}$. $C_0 = \sigma^{-3}$.  Uncertainties on the last digits are $\pm 1$ SD from seven independent estimates.\\
    }
    \label{tab:LJ}
    \begin{tabular}{cccc}
        \toprule
        $R\,[\sigma]$ & 
        $\Delta F_{R,u\to b}$ &
         $-k_{\rm B}T\log(V_{R,u} C_0)$ &
        $\Delta F_0$ \\
        \midrule
        $10$  & 1.62(5) & -7.35 & -5.73(5) \\
        $12$  & 2.14(13) & -7.87 & -5.73(13) \\
        $14$  & 2.52(9) & -8.31 & -5.79(9) \\
        \bottomrule
    \end{tabular}
\end{table}
\paragraph*{Cucurbit[7]uril/adamantanol host--guest complex.}
The left panel of Fig.~\ref{fig:cb7_results} shows a schematic of the simulation setup, in which the center-of-mass of 1-adamantanol is restrained within a cylinder of radius $R$ extending into the unbound region. Panels~(a) and~(b) report the probability histograms $p(z)$ and the free energy profiles $F(z)$ obtained at the different restraint radii $R$. The probability distributions are in excellent mutual agreement, displaying a sharp double peak around $z \approx 0$, corresponding to the bound state, and a broad, low-probability unbound region at $z\gtrsim 1$~nm. The bimodal structure of $p(z)$ in the bound state reflects the two possible orientations of the hydroxyl group of adamantanol within the CB7 cavity, with the --OH group pointing toward either receptor portal, in agreement with the results of Ref.~\cite{Vijay2025}. The corresponding free energy profiles display a deep minimum at the bound pose followed by a barrier near $z \approx 0.5$~nm, and reach a plateau at large values of $z\gtrsim 1.4$~nm, corresponding to fully solvated 1-adamantanol.
\begin{table}[!t]
\centering
\caption{
    \textbf{Binding affinity of the CB7/1-adamantanol complex.} 
    Energies in kcal/mol. $C_0 = 1$ mol/L.  Uncertainties on the last digits are $\pm 1$ SD from six independent estimates.\\
}
\label{tab:cb7_results}
\begin{tabular}{cccc}
\toprule
$R$ [nm] & $\Delta F_{R,u\to b}$ & $-k_{\rm B}T\log(V_{R,u} C_0)$ & $\Delta F_0$ \\
\midrule
\multicolumn{4}{l}{\textit{Extended-region estimator}} \\[3pt]
0.1 & -20.99(28) & 3.33 & -17.66(28) \\
0.2 & -20.19(10) & 2.50 & -17.69(10) \\
0.3 & -19.55(20) & 2.02 & -17.53(20) \\
\midrule
\multicolumn{4}{l}{\textit{Hybrid estimator}} \\[3pt]
0.1 & -22.30(27) & 4.61 & -17.68(27) \\
0.2 & -21.50(11) & 3.79 & -17.71(11) \\
0.3 & -20.85(21) & 3.30 & -17.55(21) \\
\midrule
\multicolumn{4}{l}{\textit{Single-bin estimator}} \\[3pt]
0.1 & -21.28(26) & 4.61 & -16.66(26) \\
0.2 & -20.49(12) & 3.79 & -16.70(12) \\
0.3 & -19.84(21) & 3.30 & -16.54(21) \\
\bottomrule
\end{tabular}
\end{table}

The plateau level decreases systematically as the radius $R$ increases, reflecting the larger accessible volume $V_{R,u}$ associated with wider cylindrical confining regions. This trend is correctly accounted for by the volumetric correction $-k_{\rm B}T\log(V_{R,u} C_0)$. Once this is applied, the resulting standard binding free energies $\Delta F_0$ agree within statistical uncertainty across all three cylindrical radii, with a maximum spread of $0.2$~kcal/mol. This is demonstrated in Table~\ref{tab:cb7_results}, which collects results obtained using three estimators~\footnote{To be compared with the experimental value of $\Delta F_0=-14.1\pm0.2$ kcal/mol~\cite{Grimm2022}. The discrepancy with the predictions can be attributed to force-field inaccuracy~\cite{Vijay2025} .}. The extended-region and hybrid estimators are in excellent mutual agreement. The single-bin estimator, which is not a valid estimator of $K$ (Section~\ref{sec:pmf}), deviates by $\approx1$~kcal/mol. This discrepancy demonstrates that the choice of estimator deserves careful consideration.

\begin{figure}[!t]
    \centering
    \includegraphics[width=0.8\textwidth]{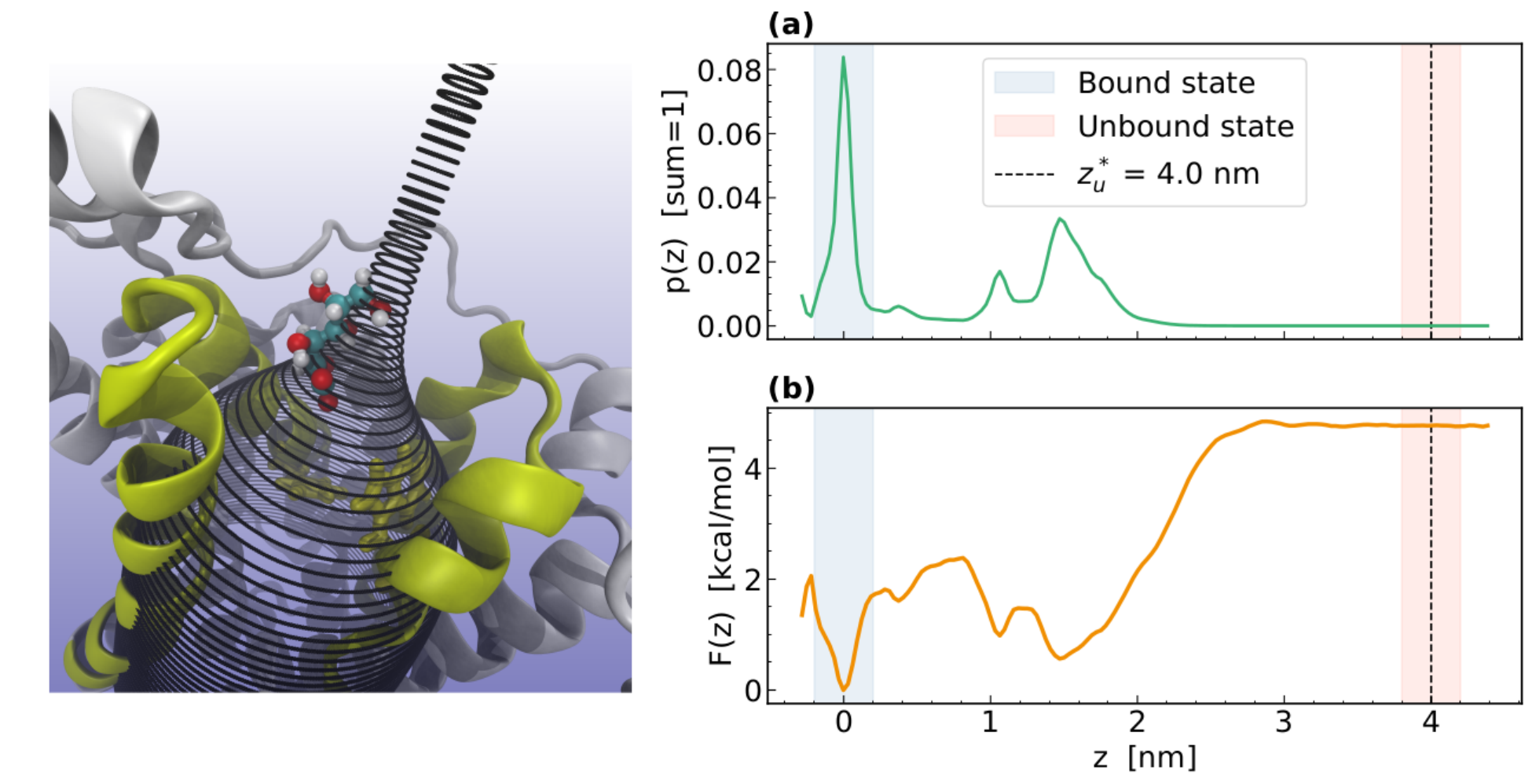}
    \caption{
    \textit{(Left)} A representative snapshot of the simulation system, showing the ligand GAL (in ball-and-stick representation, with atoms colored by element: carbon cyan, oxygen red and hydrogen white) inside the smooth confining potential (black lines) positioned along the binding $z$-axis of the transporter DgoT (yellow helices representing the transmembrane domain and gray representing the remainder of the protein).
    \textit{(Right)}~(a) Probability histogram $p(z)$. (b) Free energy profile $F(z) = -k_{\rm B}T \log p(z)$, shifted so that the global minimum is zero. The blue and red shaded regions indicate the bound and unbound integration windows used for the extended-region estimator of $\Delta F_0$. The dashed vertical line marks the single-bin unbound reference state $z_u^*=4.0$~nm.}
    \label{fig:dgot_results}
\end{figure}
\paragraph*{Galactonate--DgoT ligand--protein complex.}
The left panel of Fig.~\ref{fig:dgot_results} shows a schematics of the simulation setting, in which a smooth confining potential was used to guide GAL along the unbinding $z$-axis. The probability histogram $p(z)$ and the corresponding free energy profile $F(z)$ are shown in panels (a) and (b), respectively. The sharp peak of $p(z)$ near $z \approx 0$ identifies the primary binding pose of the substrate deep inside the transport canal, in agreement with the crystallographic binding site of the substrate~\cite{Leano2019}. The two secondary metastable minima at intermediate $z$ values reflect transient interactions of GAL with gating residues along the release pathway~\cite{Plate2026}. The unbound state is reached at values of $z \gtrsim 3.5$~nm. Accordingly, the free energy profile $F(z)$ rises monotonically beyond $z \approx 1.5$~nm, reaching a well-defined plateau for $z \gtrsim 3.5$~nm. This confirms that GAL is fully solvated in this region and no longer interacts with the protein.
\begin{table}[!t]
\centering
\caption{
    \textbf{Binding affinity of the GAL/DgoT ligand-protein complex} 
    Energies in kcal/mol. $C_0 = 1$ mol/L. Uncertainties on the last digit are maximum deviations from the best estimate over the last 4 $\mu$s of simulation.\\
}
\label{tab:dgot_results}
\begin{tabular}{cccc}
\toprule
 & $\Delta F_{R,u\to b}$ & $-k_{\rm B}T\log(V_{R,u} C_0)$ & $\Delta F_0$ \\
\midrule
Extended & -4.1(2) & 3.0 & -1.1(2) \\
Hybrid & -5.7(2) & 4.6 & -1.1(2) \\
Single-bin & -4.7(3) & 4.6 & -0.2(3) \\
\midrule
Extended$^*$ & -4.7(1) & 3.0 & -1.7(1) \\
\bottomrule
\end{tabular}
\end{table}

Table~\ref{tab:dgot_results} reports the values of the binding affinity from three different estimators. Similarly to the host-guest system results, while the extended-region and hybrid estimators yielded the same value of $\Delta F_0 \approx -1.1$~kcal / mol, the single-bin measure deviates by $\approx 1$~kcal / mol. The observed small magnitude of $\Delta F_0$ is physically consistent with the biological role of DgoT as a transporter rather than a tight binder for GAL~\cite{Dmitrieva2024,Plate2026}. 

The last estimate in Table~\ref{tab:dgot_results} was obtained defining the bound state as the complementary of the unbound one shown in Fig.~\ref{fig:dgot_results}, and it is discussed in the next section.
\section{Discussion: defining the bound and unbound states, and matching the experiment}
\label{sec:discussion}
Eq.~\eqref{eq.cornerstone} defines $K$ through the statistical weights of two distinct thermodynamic states. Turning that partition into a number requires fixing their boundaries. The unbound state presents no difficulty. It is characterized by the absence of interaction rather than by any internal arrangement, and can be read off from the plateau of the free-energy profile (or radial distribution function) at large separation. All the freedom resides in the definition of the bound state. That the binding constant depends on this definition has long been recognized~\cite{Jorgensen1989,DuboueDijon2021}: the dependence is weak for tight, well-localized complexes and becomes significant for weak or structured ones. In cases where the bound state is a single well-defined pose, $P_b$ is unambiguous. If instead it resolves into more sub-basins $\{b_j\}$ (distinct poses or binding modes), then $P_b=\sum_j P_{b_j}$, and, from Eq.~\eqref{eq.cornerstone}, $K=\sum_j K_j$\footnote{This relation treats the sub-basins as mutually exclusive, which holds when the probability that a receptor binds more than one ligand at the same time is negligible, as in the dilute limit considered in this work.}. The affinity then depends on which terms are included in the sum. Whether this is problematic depends on the simulation setting. In what follows, we consider only the case discussed in this work: the setting in which the binding constant is obtained from a single simulation in the presence of a volumetric restraint, as for example in funnel metadynamics~\cite{Limongelli2013}, that is, from the estimators of Sections~\ref{sec:restraints} and~\ref{sec:practical} with $\Delta F_{\rm corr}=0$.

These estimators require the restraining potential to be inactive whenever the ligand occupies the bound state. This is often not an issue: the confining potential can be engineered to encompass only the bulk solvent region, with its boundary placed far enough from the binding site that the ligand cannot reach it while bound. In other cases, however, the accessible region may not be so cleanly separated from the binding site. We refer to the former as the full-access setting, and to the latter as the partial-access setting. In a full-access setting, the entire bound state is sampled under the true interaction potential: every $P_{b_j}$ is reliable, and $P_b$ -- or any partial sum -- is under control. In a partial-access setting, the confining potential is inactive in some sub-basins but exerts a configuration-dependent force in others. The weights of the latter are altered by the restraint, and $P_b$ can be trusted only if the count is restricted to the ``untainted'' sub-basins, or if the ``tainted'' weight is negligible. In the former case, the binding constant so obtained is a pose-specific (or binding-mode-specific) quantity, characterizing only the untainted sub-basin(s) rather than the full bound state.

The three systems studied here span these situations. The LJ dimer uses a full-access spherical restraint and has a single dominant pose, so the affinity depends only weakly on the boundary: moving it from $2.5\,\sigma$ to $1.8\,\sigma$ shifts $\Delta F_0$ by $\approx 1.4\%$ (see Section~\ref{subsec:LJ} of SI). The CB7/1-adamantanol complex uses a partial-access setting, but a single basin dominates and the restraint is inactive there. As a consequence, including or excluding the tainted region leaves the estimate unchanged (see Section~\ref{subsec:cb7} of SI). The galactonate--DgoT complex has a structured bound state, and the restraint leaves the main pose untouched while possibly affecting the secondary sub-basins (see Section~\ref{subsec:dgot} of SI). In this case, including the tainted region shifts the prediction by $\approx -0.6$~kcal/mol (see the first and last estimates reported in Table~\ref{tab:dgot_results}).

We stress that none of the above considerations is a limitation of the theory. Population-based affinities require the partition to be specified. The present formalism makes all the choices explicit and, in full-access settings, controllable.

Which of these choices is admissible depends on what the experiment resolves~\cite{Renaud2016,DuboueDijon2021}. From this point of view, the available experimental techniques fall into two broad classes. (i) Bulk equilibrium methods~\cite{Myszka1999,VelazquezCampoy2004,Huber2006,Niesen2007,Chaires2008,Ladbury2010,Seidel2013} and separation-based assays~\cite{Wong1993,Annis2007,Hellman2007,Ryder2008,Pollard2010,Jarmoskaite2020} that do not access structural information. (ii) Site-resolved measurements, which can also provide it~\cite{LinRiggs1972,Pellecchia2008,Williamson2013,Cala2014}. These include protein-observed NMR, where the chemical-shift perturbations of assigned resonances identify the residues contacted by the ligand, while the titration assay yields the affinity, in principle even a distinct value for each well-resolved binding mode~\footnote{Chemical shifts are population-weighted averages under fast exchange, so sub-basins whose contact patterns overlap are reported as a single apparent state. Resolving them requires that exchange be slow on the chemical-shift timescale~\cite{Williamson2013}.}.

For the large majority of reported affinities, the experimental counterpart is the inclusive bound state, and a faithful comparison requires using a full-access setting. On the other hand, when the experiment does select a sub-basin, the simulation setting must leave precisely that sub-basin untainted. In this case a restricted definition is the correct target. If a partial-access setting is used for predictions and $P_b$ does depend on whether the tainted region is included, the result is a conditional prediction, valid under the hypothesis that the measured state coincides with the sub-basin targeted in the simulation~\footnote{We are not concerned here with errors arising from the choice of the model Hamiltonian. Given a sufficiently accurate model and a controlled experiment, the access-related aspects discussed in this section are the only source of discrepancy within the theory, aside from the explicit assumptions entering its derivation. With an inaccurate model, however, they may become secondary to the errors incurred elsewhere.}.

\section{Conclusions}
We have presented a statistical-mechanical treatment of equilibrium binding constants computed from molecular simulations along the geometrical route, with a focus on simulations that employ volumetric restraints. The treatment is restricted to the high dilution limit, neglecting corrections at finite concentration~\cite{deJong2011}. Our starting point is the relation between $K$ and the statistical weights of the bound and unbound states, Eq.~\eqref{eq.cornerstone}, which we re-derived from the canonical partition function of molecular solutes in explicit solvent.

For a restraint that leaves the bound state unaffected and confines a homogeneous unbound region, the system volume entering the volumetric term is replaced by $V_{R,u}$, the volume of the region over which the unbound population is evaluated. This is consistent with the results established for unrestrained simulations~\cite{deJong2011,JostLopez2020}. If the restraint also acts on the bound state, the same relation holds with an additional correction term, $\Delta F_{\rm corr}$, which is not available from the restrained simulation alone and requires separate free-energy calculations or reweighting~\cite{Woo2005,Deng2009,DuboueDijon2021}.

From the relation in Eq.~\eqref{eq.cornerstone}, we derived practical estimators that can be applied directly to unbiased or enhanced-sampling molecular dynamics simulations, requiring only histograms of the configurations sampled along a suitably defined reaction coordinate. The unbound state is identified by a coordinate describing the relative position of the two molecules, while the bound state can be defined according to whatever criterion is appropriate for the system at hand. The PMF-based expressions previously reported in previous literature~\cite{Jorgensen1989,Allen2004,Woo2005,deJong2011,Limongelli2013,DuboueDijon2021} are recovered as the special case in which the bound state is also defined along the same one-dimensional coordinate.

Because every term is derived explicitly, these expressions allowed us to identify inconsistencies in previous works. First, a single-bin estimator repeatedly used in the literature~\cite{Sun2014,Comitani2016,Zlobin2021,Szlenk2021,Kuzovlev2023,Borowska2024}, in which the bound state is reduced to the bin at the absolute minimum of the free energy profile, is not a valid estimator of $K$, since its result scales with the bin width and systematically underestimates the affinity. In both molecular complexes studied here, we find that it departs from the correct estimators by $\approx 1$~kcal/mol, comparable to the accuracy typically sought in comparisons with experiment~\cite{Ross2023}. Second, we caution against previously published working formulas that omit the bin size entering the volumetric term~\cite{Limongelli2013,Souza2020,Lamprakis2021}, which are dimensionally inconsistent as printed and yield results that depend on the unit of length.

Taken together, these features make the present treatment a self-contained practical reference, grounded in first principles, for computing binding affinities using molecular simulations.


\section*{Supporting Information}
The Supporting Information is appended at the end of this document.

Complete derivation of the central relation and of the volumetric term for
restrained simulations, simulation and analysis protocols, convergence tests,
and binding affinities obtained under alternative definitions of the bound and
unbound states.

\section*{Acknowledgments}
D.M. acknowledges support by the European Union’s HORIZON MSCA Doctoral Networks program, under Grant Agreement No. 101072344, project AQTIVATE (Advanced computing, QuanTum algorIthms and datadriVen Approaches for Science, Technology, and Engineering). D.M. gratefully acknowledges discussions with Paolo Carloni. E.I. wishes to thank Riccardo Capelli for valuable discussions. D.M. and E.I. thank Amal Vijay for providing templates for the host-guest system simulations.


\bibliographystyle{unsrtnat}
\bibliography{refs}

\clearpage
\FloatBarrier

\begin{center}
  {\LARGE\bfseries Supporting Information}\\[1.2ex]
  {\large From populations to absolute binding constants in molecular simulations:
   exact volumetric terms and practical estimators}\\[1.2ex]
  {\normalsize Davide Mandelli, Emiliano Ippoliti, and Charles Plate}
\end{center}
\vspace{1em}

\setcounter{section}{0}
\setcounter{figure}{0}
\setcounter{table}{0}
\setcounter{equation}{0}
\renewcommand{\thesection}{S\arabic{section}}
\renewcommand{\thesubsection}{S\arabic{section}.\arabic{subsection}}
\renewcommand{\thesubsubsection}{S\arabic{section}.\arabic{subsection}.\arabic{subsubsection}}
\renewcommand{\theequation}{S\arabic{equation}}
\renewcommand{\thefigure}{S\arabic{figure}}
\renewcommand{\thetable}{S\arabic{table}}
\renewcommand{\theHsection}{S\arabic{section}}
\renewcommand{\theHequation}{S\arabic{equation}}
\renewcommand{\theHfigure}{S\arabic{figure}}
\renewcommand{\theHtable}{S\arabic{table}}


\section{Theory}

\subsection{Statistical mechanics of chemical equilibrium}
\label{subsec:statmec}

\subsubsection*{The partition function}
Related derivations of the relation obtained at the end of this section have been given starting from a single receptor in a dilute solution of ligands~\cite{Roux1996,Woo2005} and, for structureless particles, from the configuration integral of an arbitrary number of dimerizing particles~\cite{deJong2011}. Here we consider molecular solutes in explicit solvent. Accordingly, we start our considering the canonical ensemble of a system containing $N_A$ molecules of species $A$, $N_B$ molecules of species $B$, and $N_S$ solvent molecules. The system is enclosed in a macroscopic volume $V$ and coupled to a heat reservoir at temperature $T$. Without loss of generality, we assume $N_A \leq N_B$. We denote by $M_\alpha$ the number of atoms composing a molecule of species $\alpha \in \{A, B, S\}$, so that the total number of atoms is $M = M_A N_A + M_B N_B + M_S N_S$. For convenience we also define $M_{AB}=M_A+M_B$ as the total number of atoms composing a $AB$ complex.

We assume that molecules of the reactive species $A$ and $B$ can dynamically
associate into a complex $AB$ or dissociate from it
\begin{equation}
    A + B \rightleftharpoons AB.
\end{equation}
At any instant of time, the system will contain a certain number $N_{AB}$ of bound pairs together with $N_A^u = N_A - N_{AB}$ and $N_B^u = N_B - N_{AB}$ unbound molecules of species $A$ and $B$, respectively. Our goal is to identify the statistically dominant state $\{\bar{N}_A^u, \bar{N}_B^u, \bar{N}_{AB}\}$, which yields the equilibrium concentrations $[A]=\bar{N}_A^u/V$, $[B]=\bar{N}_B^u/V$, and $[AB]=\bar{N}_{AB}/V$.

We collect all degrees of freedom of the $i$th molecule of species $A$, $B$, $S$ into the vectors $\mathbf{a}_i$, $\mathbf{b}_i$, $\mathbf{s}_i$, respectively, and denote by $\mathbf{ab}_i$ all degrees of freedom of the $i$th bound pair. We further write $\mathbf{A}$, $\mathbf{B}$, $\mathbf{S}$, and $\mathbf{AB}$ for the collective degrees of freedom of \emph{all} molecules of the corresponding species. With this notation the canonical partition function, Eq.~\eqref{eq.Z} of the main text, reads
\begin{equation}
\label{eq:Q}
    \mathcal{Z} = \frac{1}{M!\, h^{3M}}
    \int \mathrm{d}\mathbf{A}\, \mathrm{d}\mathbf{B}\, \mathrm{d}\mathbf{S}
    \; e^{-\beta \mathcal{H}},
\end{equation}
where $\beta = (k_{\rm B} T)^{-1}$, $\mathcal{H}$ is the Hamiltonian of the system, and the prefactor $(M!\,h^{3M})^{-1}$ is the standard correction for particle indistinguishability and phase space dimensionality. $h$ is Planck's constant.

For a given value $N_{AB} \in \{0, 1, \ldots, N_A\}$, the number of distinct ways of selecting $N_{AB}$ molecules $A$ and $N_{AB}$ molecules $B$ to form $N_{AB}$ labelled pairs is
\begin{equation}
    \binom{N_A}{N_{AB}}\binom{N_B}{N_{AB}} N_{AB}! =
    \frac{N_A!\, N_B!}{N_A^u!\, N_B^u!\, N_{AB}!}.
\end{equation}
All such realizations contribute identically to the partition function. Summing over all possible values of $N_{AB}$ and accounting for this combinatorial degeneracy, we can rewrite
\begin{equation}
\label{eq:Q_split}
    \mathcal{Z} = \frac{1}{M!\,h^{3M}}
    \sum_{N_{AB}=0}^{N_A}
    \frac{N_A!\,N_B!}{N_A^u!\,N_B^u!\,N_{AB}!}
    \int \mathrm{d}\mathbf{S}
    \int_u \mathrm{d}\mathbf{A}_u
    \int_u \mathrm{d}\mathbf{B}_u
    \int_b \mathrm{d}\mathbf{AB}\;
    e^{-\beta \mathcal{H}}.
\end{equation}
Here $\mathbf{A}_u$ and $\mathbf{B}_u$ denote the collective degrees of freedom of the \textit{selected} $N_A^u$ and $N_B^u$ unbound molecules, $\mathbf{AB}$ denote the collective degrees of freedom of the \textit{selected} $N_{AB}$ bound pairs. The \textit{restricted} integrals $\int_u \mathrm{d}\mathbf{A}_u$ and $\int_u \mathrm{d}\mathbf{B}_u$ run over the subregion of phase space corresponding to unbound configurations of the selected molecules, i.e.\ configurations in which no molecule of species $A$ is paired with any molecule of species $B$ to form a complex. Conversely, $\int_b \mathrm{d}\mathbf{AB}$ is restricted to bound configurations of the selected $N_{AB}$ pairs, i.e.\ configurations in which each selected molecule of $A$ is associated with its paired molecule of $B$ to form a complex $AB$. The precise definition of bound and unbound configurations is deliberately left unspecified at this stage. We assume only that the two states can be unambiguously identified, and that they are mutually exclusive and collectively exhaustive~\footnote{Relaxation of this assumption is discussed in Section~\ref{sec:discussion} of the main text.}. The integral $\int\mathrm{d}\mathbf{S}$ runs over the full phase space of the solvent.

We stress that while we consider here the specific case of two species forming a single complex, the formalism extends straightforwardly to multiple species and multiple complexes, including the case in which  $A$ and $B$ can adopt multiple structurally distinct bound states (e.g., different binding modes or stoichiometries).

\subsubsection*{The dilute limit}

We now turn to the Hamiltonian $\mathcal{H}$. Assuming pairwise interactions only,
we write
\begin{eqnarray}
    \mathcal{H} &=& \sum_{i=1}^{N_A} h_A(\mathbf{a}_i)
        + \sum_{i=1}^{N_B} h_B(\mathbf{b}_i)
        + \sum_{i=1}^{N_S} h_S(\mathbf{s}_i) \nonumber\\
      &+& \sum_{\langle ij\rangle} h_{AA}(\mathbf{a}_i,\mathbf{a}_j)
        + \sum_{\langle ij\rangle} h_{BB}(\mathbf{b}_i,\mathbf{b}_j)
        + \sum_{\langle ij\rangle} h_{SS}(\mathbf{s}_i,\mathbf{s}_j) \nonumber\\
      &+& \sum_{\langle ij\rangle} h_{AB}(\mathbf{a}_i,\mathbf{b}_j)
        + \sum_{\langle ij\rangle} h_{AS}(\mathbf{a}_i,\mathbf{s}_j)
        + \sum_{\langle ij\rangle} h_{BS}(\mathbf{b}_i,\mathbf{s}_j),
\end{eqnarray}
where $h_\alpha(\boldsymbol{\alpha}_i)$ collects the kinetic energy and intramolecular interactions of the $i$th molecule of species $\alpha$, and $h_{\alpha\beta}(\boldsymbol{\alpha}_i,\boldsymbol{\beta}_j)$ denotes the intermolecular interaction between the $i$th
molecule of species $\alpha$ and the $j$th molecule of species $\beta$. The last six sums run over all distinct pairs $\langle ij\rangle$.

In the limit of ideal dilution of $A$ and $B$, the phase-space volume of configurations in which two or more molecules of the same species interact is negligible, hence, we can set $h_{AA}=h_{BB} \approx 0$. Furthermore, for the same reason, interactions between $A$ and $B$ molecules are non-negligible only for pairs involved in a complex $AB$, which restricts the summation over $h_{AB}$ only over the selected \textit{bound} pairs. Finally, in the dilute limit, the solvent-solute interactions $h_{AS}$ and $h_{BS}$ become essentially independent of the specific positions of the $A$ and $B$ molecules and depend only on the solvent configuration. The effective Hamiltonian then decouples as
\begin{equation}
    \mathcal{H} \approx \mathcal{H}_S + \mathcal{H}_{AB},
\end{equation}
where
\begin{equation}
    \mathcal{H}_S = \sum_{i=1}^{N_S} h_S(\mathbf{s}_i)
        + \sum_{\langle ij\rangle} h_{SS}(\mathbf{s}_i,\mathbf{s}_j)
        + \sum_{\langle ij\rangle} h_{AS}(\mathbf{a}_i,\mathbf{s}_j)
        + \sum_{\langle ij\rangle} h_{BS}(\mathbf{b}_i,\mathbf{s}_j)
\end{equation}
contains all solvent degrees of freedom and solvent--solute interactions~\footnote{As mentioned above, in this limit, $h_{BS}$ and $h_{AS}$ formally do not depend on $\{{\bf a}_i,{\bf b}_i\}$. This is roughly equivalent to assuming the solute molecules as rigid.}, while
\begin{equation}
    \mathcal{H}_{AB} = \sum_{i=1}^{N_A} h_A(\mathbf{a}_i)
           + \sum_{i=1}^{N_B} h_B(\mathbf{b}_i)
           + \sum_{\langle ij\rangle_{\rm bound}} h_{AB}(\mathbf{a}_i,\mathbf{b}_j)
\end{equation}
depends on the reactive species only.
The partition function Eq.~\eqref{eq:Q_split} factorises as
\begin{eqnarray}
\label{eq:Q_factored}
    \mathcal{Z} &=& C\, Z_S \sum_{N_{AB}=0}^{N_A}
    \frac{1}{N_A^u!}\left(\int_u \mathrm{d}\mathbf{a}\, \frac{e^{-\beta h_A(\mathbf{a})}}{h^{3M_A}}\right)^{N_A^u}\times\nonumber\\
    &&\frac{1}{N_B^u!}\left(\int_u \mathrm{d}\mathbf{b}\, \frac{e^{-\beta h_B(\mathbf{b})}}{h^{3M_B}}\right)^{N_B^u}\times\nonumber \\
    &&\frac{1}{N_{AB}!}\left(\int_b \mathrm{d}\mathbf{ab}\, \frac{e^{-\beta h_{AB}(\mathbf{ab})}}{h^{3M_{AB}}}\right)^{N_{AB}},
\end{eqnarray}
where the combinatorial prefactor is
\begin{equation}
    C = \frac{N_A!\, N_B!\, N_S!}{M!},
\end{equation}
and the solvent partition function is
\begin{equation}
\label{eq:ZS}
    Z_S = \frac{1}{h^{3M_SN_S}N_S!}
    \int \mathrm{d}\mathbf{S}\; e^{-\beta \mathcal{H}_S}.
\end{equation}
In Eq.~\eqref{eq:Q_factored}, the integrals ($\int_u$) over a single unbound molecule run over
the full single-molecule phase space, since in the dilute limit the
volume of bound configurations is negligible relative to the rest, and
the restriction to unbound configurations can be lifted without
appreciable error. The integral $\int_b
\mathrm{d}\mathbf{ab}$ is restricted to bound configurations of a
single $AB$ pair. 

Introducing the single-molecule partition functions
\begin{equation}
\label{eq:z}
    z_A = \int \mathrm{d}\mathbf{a}\; \frac{e^{-\beta h_A(\mathbf{a})}}{h^{3M_A}},
    \quad
    z_B = \int \mathrm{d}\mathbf{b}\; \frac{e^{-\beta h_B(\mathbf{b})}}{h^{3M_B}},
    \quad
    z_{AB} = \int_b \mathrm{d}\mathbf{ab}\;\frac{e^{-\beta h_{AB}(\mathbf{ab})}}{h^{3M_{AB}}} ,
\end{equation}
equation Eq.~\eqref{eq:Q_factored} takes the compact form
\begin{eqnarray}
\label{eq:Q_compact}
    \mathcal{Z} &=& C\,Z_S \sum_{N_{AB}=0}^{N_A}
    \frac{(z_A)^{N_A^u}}{N_A^u!}
    \frac{(z_B)^{N_B^u}}{N_B^u!}
    \frac{(z_{AB})^{N_{AB}}}{N_{AB}!}\nonumber\\
    &=& C\,Z_S \sum_{N_{AB}=0}^{N_A} \Omega(N_A^u, N_B^u, N_{AB}),
\end{eqnarray}

\subsubsection*{Thermodynamic limit and equilibrium condition}

The Helmholtz free energy of the system is defined as
\begin{equation}
    F = -k_{\rm B}T\log\mathcal{Z}.
\end{equation}
We now show that in the thermodynamic limit only one term in the sum
of  Eq.~\eqref{eq:Q_compact} dominates. Since all summands are strictly
positive, there exists a largest one, $\Omega_{\rm max}$.
The following chain of inequalities then hold~\footnote{Strictly speaking, the sum contains $N_A+1$ summands, so the prefactor in the rightmost expression should read $N_A+1$. Since $N_A\gg1$, we drop the 1.}:
\begin{equation}
\label{eq:ineq1}
    \log\!\left(C\,Z_S\,\Omega_{\rm max}\right)
    < \log\mathcal{Z} <
    \log\!\left(N_A\,C\,Z_S\,\Omega_{\rm max}\right).
\end{equation}
In the thermodynamic limit ($N_\alpha,V\to+\infty$ but keeping constant concentrations $N_\alpha/V$), we can use Stirling's approximation:
\begin{equation}
    \log(n!) = n\log n - n + O(\log n)\xrightarrow{n\to+\infty} n\log n - n
\end{equation}
to show that 
\begin{equation}
    \log C = \log\left(\frac{N_A!\,N_B!\,N_S!}{M!}\right) \approx
M\sum_{\alpha}\frac{N_\alpha}{M}\log \frac{N_\alpha}{M} \propto M \gg \log N_A, 
\end{equation}
because the fraction of each species is kept fixed in the thermodynamic limit, and therefore
\begin{equation}
\frac{\log N_A}{\log(C\,Z_S\,\Omega_{\rm max})} = \frac{\log N_A}{\log C +\log(Z_S\,\Omega_{\rm max})} < \frac{\log N_A}{\log C} \xrightarrow{N_\alpha\to+\infty}  0.
\end{equation}
This implies that  Eq.~\eqref{eq:ineq1} can be rewritten as
\begin{equation}
    1    
    < \frac{\log\mathcal{Z}}{\log\!\left(C\,Z_S\,\Omega_{\rm max}\right)} 
    < \frac{\log N_A}{\log\left(C\,Z_S\,\Omega_{\rm max}\right)} + 1
    < \frac{\log N_A}{\log C} + 1
    \xrightarrow{N_\alpha\to+\infty}  1,
\end{equation}
or
\begin{equation}
   \log\mathcal{Z} \to \log\!\left(C\,Z_S\,\Omega_{\rm max}\right).
\end{equation}
Consequently, in the thermodynamic limit, the Helmholtz free energy is
\begin{equation}
    F = -k_{\rm B}T\log\!\left[C\,Z_S\,
    \Omega(\bar{N}_A^u,\bar{N}_B^u,\bar{N}_{AB})\right],
\end{equation}
where $\{\bar{N}_{A}^u,\bar{N}_B^u,\bar{N}_{AB}\}$ maximize
$\Omega$.

To find the dominant term in $\Omega$ we impose the stationary condition $\delta\Omega = 0$ under the
constraint that $\delta N^u_A = \delta N^u_B = -\delta N_{AB}$
\begin{eqnarray*}
\delta\Omega (N_A^u,N_B^u,N_{AB}) &=& \frac{\partial \Omega}{\partial N_A^u} \delta N_A^u + \frac{\partial \Omega}{\partial N_B^u} \delta N_B^u + \frac{\partial \Omega}{\partial N_{AB}} \delta N_{AB} = \\
&=& \left( \frac{\partial \Omega}{\partial N_A^u} + \frac{\partial \Omega}{\partial N_B^u} - \frac{\partial \Omega}{\partial N_{AB}} \right)  \delta N_A^u = 0 
\end{eqnarray*}
which implies
\begin{equation}
\label{eq:omega_equil}
\frac{\partial \Omega}{\partial N_A^u} + \frac{\partial \Omega}{\partial N_B^u} = \frac{\partial \Omega}{\partial N_{AB}}.
\end{equation}
Applying the mathematical identity:
\begin{equation*}
\frac{\partial \Omega}{\partial x} = \Omega \cdot \frac{\partial \log \Omega}{\partial x},  
\end{equation*}
we can rewrite Eq.~\eqref{eq:omega_equil} as
\begin{equation}
    \Omega \cdot \left[\frac{\partial \log \Omega}{\partial N_A^u} \right] +
    \Omega \cdot \left[\frac{\partial \log \Omega}{\partial N_B^u} \right] =
    \Omega \cdot \left[\frac{\partial \log \Omega}{\partial N_{AB}} \right] 
\end{equation}
and since $\Omega > 0$:
\begin{equation}
    \left[\frac{\partial \log \Omega}{\partial N_A^u} \right] +
    \left[\frac{\partial \log \Omega}{\partial N_B^u} \right] =
    \left[\frac{\partial \log \Omega}{\partial N_{AB}} \right].
\end{equation}
Using the definition of $\Omega$ in  Eq.~\eqref{eq:Q_compact}, the last equation becomes
\begin{equation}
\label{eq:equil}
    \frac{\partial}{\partial N_A}\log\!\frac{(z_A)^{N_A}}{N_A!}
    +\frac{\partial}{\partial N_B}\log\!\frac{(z_B)^{N_B}}{N_B!}
    = \frac{\partial}{\partial N_{AB}}\log\!\frac{(z_{AB})^{N_{AB}}}{N_{AB}!},
\end{equation}
where we have omitted the superscript $u$ on the unbound counts for brevity.
Defining the free energy and chemical potential of species $\alpha$
at concentration $[\alpha] = N_\alpha/V$ as
\begin{equation}
    F_\alpha = -k_{\rm B}T\log\!\frac{(z_\alpha)^{N_\alpha}}{N_\alpha!},
    \qquad
    \mu_\alpha = \frac{\partial F_\alpha}{\partial N_\alpha},
\end{equation}
 Eq.~\eqref{eq:equil} yields the expected equilibrium condition in terms of the chemical potentials
\begin{equation}
    \mu_A + \mu_B = \mu_{AB}.
\end{equation}

The terms inside the derivatives in  Eq.~\eqref{eq:equil} can be simplified by using Stirling's formula:
\begin{equation*}
\log\frac{z^N}{N!} = N\log z - \log N! \approx 
N\log z - (N\log N - N) = 
N\log z - N\log N + N,
\end{equation*}
and their derivative with respect to $N$ can be approximated as
\begin{equation}
\frac{\partial }{\partial N} \log\frac{z^N}{N!} \approx \log z - \left( \log N + N\cdot\frac{1}{N} \right) + 1 = \log z - \log N = \log \frac{z}{N},
\end{equation}
yielding
\begin{equation}
\label{eq:Kd_z}
    \frac{z_{AB}}{z_A z_B}
    = \frac{[AB]}{[A][B]}\,\frac{1}{V}.
\end{equation}
This equation connects the ratio of single-molecule partition functions to the equilibrium constant $K=\frac{[AB]}{[A][B]}$.

We can explicitly integrate over the centre-of-mass
coordinates and momenta in $z_\alpha$. Separating centre-of-mass
and internal degrees of freedom gives
\begin{equation}
\label{eq:z_alpha}
    z_\alpha = \frac{V}{\Lambda_\alpha^3}\,z_\alpha^{\rm int},
\end{equation}
where
\begin{equation}
    \Lambda_\alpha = \frac{h}{\sqrt{2\pi m_\alpha k_{\rm B}T}}
\end{equation}
is the thermal de~Broglie wavelength of species $\alpha$ with total
molecular mass $m_\alpha$, and
\begin{equation}
    z_\alpha^{\rm int} =
    \int \mathrm{d}\boldsymbol{\xi}_\alpha\,
    \frac{e^{-\beta h_\alpha^{\rm int}}}{h^{3(M_\alpha-1)}}
\end{equation}
is the internal partition function, with $h_\alpha^{\rm int}$
the intramolecular Hamiltonian and $\boldsymbol{\xi}_\alpha$ representing all the
internal degrees of freedom over which we have not integrated yet. Substituting  Eq.~\eqref{eq:z_alpha} into  Eq.~\eqref{eq:Kd_z}
and introducing a standard concentration $C_0$ to keep the
arguments of the logarithms dimensionless, we obtain
\begin{equation}
\label{eq:DG0_zint}
    \Delta F_0 = -k_{\rm B}T\log\!\left(
    \frac{z_{AB}^{\rm int}}{z_A^{\rm int}\,z_B^{\rm int}}
    \frac{\Lambda_A^3\,\Lambda_B^3}{\Lambda_{AB}^3}\,C_0
    \right),
\end{equation}
where we have defined the standard binding free energy
\begin{equation}
    \Delta F_0 = -k_{\rm B}T\log(K\,C_0).
\end{equation}
The interpretation of $\Delta F_0$ as a standard free energy difference follows from the chemical potential at concentration $C_0$,\footnote{The chemical potential of species $\alpha$ at arbitrary concentration
$[\alpha] = N_\alpha/V$ follows from the definition $\mu_\alpha = \partial F_\alpha/\partial N_\alpha$ with
$F_\alpha = -k_{\rm B}T\log\left[(z_\alpha)^{N_\alpha}/N_\alpha!\right]$. Applying Stirling's approximation,
\begin{equation*}
    \mu_\alpha
    \approx -k_{\rm B}T\frac{\partial}{\partial N_\alpha}
      \bigl(N_\alpha\log z_\alpha - N_\alpha\log N_\alpha + N_\alpha\bigr)
    = -k_{\rm B}T\log\frac{z_\alpha}{N_\alpha}
    = -k_{\rm B}T\log\frac{z_\alpha^{\rm int}\,V}{\Lambda_\alpha^3\,N_\alpha}.
\end{equation*}
Setting $N_\alpha/V = C_0$ gives the standard chemical potential
\begin{equation*}
    \mu_{0}^{\alpha}
    = -k_{\rm B}T\log\frac{z_\alpha^{\rm int}}{\Lambda_\alpha^3\,C_0}.
\end{equation*}}
\begin{equation}
    \mu_{0}^{\alpha} = -k_{\rm B}T\log\!\left(
    \frac{z_\alpha^{\rm int}}{\Lambda_\alpha^3\,C_0}\right),
\end{equation}
which gives immediately
\begin{equation}
    \Delta F_0 = \mu_{0}^{AB} - \mu_{0}^{A} - \mu_{0}^{B}.
\end{equation}
In other words, $\Delta F_0$ is the reversible work required to remove one molecule of $A$ and one molecule of $B$ from solutions at concentration $C_0$ and form one complex $AB$ in a solution at the same concentration.\footnote{By definition of chemical potential, $\delta F^\alpha = \mu^\alpha\,\delta N^\alpha$, so $\delta F_0^\alpha = \pm\mu_0^\alpha$ is the free energy change for adding or removing one molecule at standard concentration.} This interpretation does not rely on the absence of an explicit solvent: in the dilute limit, $-k_{\rm B}T\log(KC_0)$ is the difference between the standard chemical potential of the complex and those of its separated partners also when these include the interaction with the solvent~\cite{deJong2011}. It therefore applies equally to the explicit-solvent result derived below, Eq.~\eqref{cs}.

In principle, Eq.~\eqref{eq:DG0_zint} is  amenable to evaluation via molecular simulations as the internal partition functions can be calculated using simulations of isolated molecules \textit{in vacuo}. In practice, however, this approach is limited to small molecules for which solvent effects are negligible, since the derivation rests on the ideal dilution approximation in which solvent--solute interactions enter only through $\mathcal{H}^S$ and do not explicitly modulate the $A$--$B$ interaction. An implicit-solvent treatment is possible within a
mean-field description of the solvent, but is generally inadequate for realistic biomolecular systems, where the stability and structure of macromolecules such as proteins depend critically on explicit interactions with the solvent environment.  As we show in the next section, an equivalent expression can be derived that maps directly onto the framework of explicit-solvent molecular simulations.

\subsubsection*{Derivation with explicit solvent}

Returning to  Eq.~\eqref{eq:Q_split}, we now explicitly retain the solvent degrees of freedom throughout. We label the $N_{AB}$ molecules forming a complex, such that the $i$th molecule of species $A$ is paired with the $i$th molecule of species $B$ in the complex, while the remaining $(N_{A} - N_{AB})$ molecules of species $A$ and $(N_B - N_{AB})$ molecules of species $B$ are unbound.
Under the pairwise interaction assumption, the Hamiltonian reads
\begin{eqnarray}
    \mathcal{H} &=& \sum_{i=1}^{N_{AB}}
          \left[h_A(\mathbf{a}_i) + h_B(\mathbf{b}_i)
                + h_{AB}(\mathbf{a}_i,\mathbf{b}_i)
               + h_{AS}(\mathbf{a}_i,\mathbf{S})
               + h_{BS}(\mathbf{b}_i,\mathbf{S})\right] \nonumber\\
      &+& \sum_{i=N_{AB}+1}^{N_A}
          \left[h_A(\mathbf{a}_i) + h_{AS}(\mathbf{a}_i,\mathbf{S})\right] + \sum_{i=N_{AB}+1}^{N_B}
          \left[h_B(\mathbf{b}_i)+ h_{BS}(\mathbf{b}_i,\mathbf{S})\right] \nonumber\\
      &+& \sum_{\langle ij\rangle_{uu}} h_{AB}(\mathbf{a}_i,\mathbf{b}_j)
        + \sum_{\langle ij\rangle_{ub}} h_{AB}(\mathbf{a}_i,\mathbf{b}_j)
        + \sum_{\langle ij\rangle_{bb'}} h_{AB}(\mathbf{a}_i,\mathbf{b}_j)\nonumber\\
        &+& \mathcal{H}_S(\mathbf{S}),
\end{eqnarray}
where $h_{\alpha S}(\boldsymbol{\alpha}_i, \mathbf{S})$ denotes the interaction Hamiltonian terms
of the $i$th molecule of species $\alpha$ with \emph{all} solvent
molecules. The subscripts $uu$, $ub$, and $bb'$ on the last three sums denote pairs of
molecules that are, respectively, both unbound, one bound and one unbound, and
both bound but to different complexes. $\mathcal{H}_S(\mathbf{S})$ contains all purely solvent
contributions.

In the limit of high dilution, configurations in which two distinct
solute molecules of any species interact are rare and contribute
negligibly to the partition function. We therefore set
\begin{equation}
    \sum_{\langle ij\rangle_{uu}} h_{AB} =
    \sum_{\langle ij\rangle_{ub}} h_{AB} =
    \sum_{\langle ij\rangle_{bb'}} h_{AB} \approx 0.
\end{equation}
The effective Hamiltonian then separates as
\begin{equation}
    \mathcal{H} \approx \mathcal{H}_S(\mathbf{S}) +\sum_{i=1}^{N_{AB}} h_{ABS}(\mathbf{ab}_i, \mathbf{S})
                + \sum_{i=N_{AB}+1}^{N_A}  h_{AS}(\mathbf{a}_i, \mathbf{S})
                + \sum_{i=N_{AB}+1}^{N_B}  h_{BS}(\mathbf{b}_i, \mathbf{S}).
\end{equation}
Here we have defined the single-pair Hamiltonian (including interactions of both molecules with the solvent),
\begin{equation*}
    h_{ABS}(\mathbf{ab}_i,\mathbf{S}) =
    h_A(\mathbf{a}_i) + h_B(\mathbf{b}_i)
    + h_{AB}(\mathbf{a}_i,\mathbf{b}_i)
    + h_{AS}(\mathbf{a}_i,\mathbf{S})
    + h_{BS}(\mathbf{b}_i,\mathbf{S}),
\end{equation*}
and the single-molecule Hamiltonians for a molecule of each
species interacting with the solvent,
\begin{eqnarray*}
    h_{AS}(\mathbf{a}_i,\mathbf{S}) &=& h_A(\mathbf{a}_i)
    + h_{AS}(\mathbf{a}_i,\mathbf{S}),
    \nonumber\\
    h_{BS}(\mathbf{b}_i,\mathbf{S}) &=& h_B(\mathbf{b}_i)
    + h_{BS}(\mathbf{b}_i,\mathbf{S}).
\end{eqnarray*}
The partition function becomes
\begin{eqnarray}
\label{eq:Q_solvent}
    \mathcal{Z} &\approx& C\sum_{N_{AB}=0}^{N_A}
    \int\mathrm{d}\mathbf{S}\,\frac{e^{-\beta \mathcal{H}_S(\mathbf{S})}}{h^{3M_SN_S}}
    \;\frac{1}{N_A^u!}
    \left(\int_u\mathrm{d}\mathbf{a}\,
    \frac{e^{-\beta h_{AS}(\mathbf{a},\mathbf{S})}}{h^{3M_A}}\right)^{N_A^u}
    \times \\
    &\times&
    \frac{1}{N_B^u!}
    \left(\int_u\mathrm{d}\mathbf{b}\,
    \frac{e^{-\beta h_{BS}(\mathbf{b},\mathbf{S})}}{h^{3M_B}}\right)^{N_B^u}
    \frac{1}{N_{AB}!}
    \left(\int_b\mathrm{d}\mathbf{ab}\,\frac{e^{-\beta h_{ABS}(\mathbf{ab},\mathbf{S})}}{h^{3M_{AB}}}
    \right)^{N_{AB}},\nonumber
\end{eqnarray}
where $C$ is the same constant defined before, and the decoupling of the
unbound integrals follows from the same argument as in the previous
section. The notation $\int_{u,b}$ indicates
that the integrals are restricted to unbound or bound
configurations. Introducing the solvent-configuration-dependent single-molecule
partition functions
\begin{eqnarray}
    \label{eq:z_solvent}
    z_A(\mathbf{S}) &=& \int_u\mathrm{d}\mathbf{a}\,\frac{e^{-\beta h_{AS}(\mathbf{a},\mathbf{S})}}{h^{3M_A}}
    ,
    \nonumber \\
    z_B(\mathbf{S}) &=& \int_u\mathrm{d}\mathbf{b}\,\frac{e^{-\beta h_{BS}(\mathbf{b},\mathbf{S})}}{h^{3M_B}}
    ,
     \\
    z_{AB}(\mathbf{S}) &=& \int_b\mathrm{d}\mathbf{ab}\,\frac{e^{-\beta h_{ABS}(\mathbf{ab},\mathbf{S})}}{h^{3M_{AB}}}
    , \nonumber
\end{eqnarray}
 Eq.~\eqref{eq:Q_solvent} takes the compact form
\begin{eqnarray}
\label{eq:Q_solvent_compact}
    \mathcal{Z} &\approx& C \sum_{N_{AB}=0}^{N_A}
    \int\mathrm{d}\mathbf{S}\,\frac{e^{-\beta \mathcal{H}_S(\mathbf{S})}}{h^{3M_SN_S}}\,
    \frac{\bigl(z_A(\mathbf{S})\bigr)^{N_A^u}}{N_A^u!}
    \frac{\bigl(z_B(\mathbf{S})\bigr)^{N_B^u}}{N_B^u!}
    \frac{\bigl(z_{AB}(\mathbf{S})\bigr)^{N_{AB}}}{N_{AB}!}\nonumber\\
    &=&C \sum_{N_{AB}=0}^{N_A} \Omega\left(N_A^u, N_B^u, N_{AB}\right)
\end{eqnarray}

The key difference from the previous section is that the
single-molecule partition functions $z_\alpha(\mathbf{S})$ now depend
explicitly on the solvent configuration $\mathbf{S}$, and the
equilibrium condition must be derived by averaging over the solvent,
as we discuss next. The Helmholtz free energy is defined as before,
\begin{equation}
    F = -k_{\rm B}T\log\mathcal{Z}.
\end{equation}
And the following inequalities hold:
\begin{equation}
    \log\!\left(C\,\Omega_{\rm max}\right)
    < \log\mathcal{Z} 
    <\log\!\left(N_A\,C\,\Omega_{\rm max}\right),
\end{equation}
which, in the thermodynamic limit, lead to
\begin{equation}
    F = -k_{\rm B}T\log\!\left[C\,
    \Omega(\bar{N}_A^u, \bar{N}_B^u, \bar{N}_{AB})
    \right],
\end{equation}
where $\bar{N}_A^u$, $\bar{N}_A^u$ and $\bar{N}_{AB}$ maximises $\Omega$.
Imposing $\delta\Omega = 0$ under the constraint $\delta N^u_A = \delta N^u_B = -\delta N_{AB}$, yields
\begin{equation}
\label{eq:equil_solvent}
    \left\langle
    \log\!\left(\frac{z_{AB}(\mathbf{S})}{z_A(\mathbf{S})\,z_B(\mathbf{S})}\right)
    - \log\!\left(\frac{[AB]}{[A][B]}\,\frac{1}{V}\right)
    \right\rangle_{\!\rho} = 0,
\end{equation}
where the average is taken over the solvent degrees of freedom,
\begin{equation}
    \langle\,\cdot\,\rangle_\rho =
    \int\mathrm{d}\mathbf{S}\;(\,\cdot\,)\;\rho(\mathbf{S}),
\end{equation}
with respect to the marginal probability density of the solvent
configuration:
\begin{equation}
\label{eq:rho_S}
    \rho(\mathbf{S}) =
    \frac{1}{\mathcal{Z}}\,
    \frac{e^{-\beta \mathcal{H}_S(\mathbf{S})}}{h^{3M_SN_S}}\,
    \frac{\bigl(z_A(\mathbf{S})\bigr)^{N_A^u}}{N_A^u!}\,
    \frac{\bigl(z_B(\mathbf{S})\bigr)^{N_B^u}}{N_B^u!}\,
    \frac{\bigl(z_{AB}(\mathbf{S})\bigr)^{N_{AB}}}{N_{AB}!}.
\end{equation}
This is the probability of finding the solvent in configuration
$\mathbf{S}$, irrespective of the positions of all solute molecules.
Equation~\eqref{eq:equil_solvent} is the generalisation of
 Eq.~\eqref{eq:Kd_z} to the case of explicit solvent: the ratio of
equilibrium concentrations is now related to a solvent-averaged ratio
of partition functions, rather than to their ratio evaluated \textit{in vacuo}.

In the ideal-dilution regime, the solute molecules (whether unbound
$A$ and $B$ molecules or $AB$ complexes) are all well separated from one
another and distributed essentially homogeneously throughout the volume
$V$. We assume that, from the perspective of the solvent, all such solute configurations are equivalent and that, for each of them, the solvent configurations $\tilde{S}$ that contribute non-negligibly to the average $\langle\,\cdot\,\rangle_\rho$ contribute equally\footnote{This is roughly equivalent to considering the solute molecules rigid.}. Under this assumption, the integrand in Eq.~\eqref{eq:equil_solvent} is constant over the support of $\rho$, and the equilibrium condition reduces to a point-wise identity,
\begin{equation}
\label{eq:equil_pointwise}
    \log\!\left(
    \frac{z_{AB}(\tilde{\mathbf{S}})}
         {z_A(\tilde{\mathbf{S}})\,z_B(\tilde{\mathbf{S}})}
    \right)
    = \log\!\left(\frac{[AB]}{[A][B]}\,\frac{1}{V}\right)
    \qquad \forall\,\tilde{\mathbf{S}},
\end{equation}
or
\begin{equation}
\label{eq:equil_pointwiseB}
    \frac{z_{AB}(\tilde{\mathbf{S}})}{z_A(\tilde{\mathbf{S}})\,z_B(\tilde{\mathbf{S}})}
    = \frac{[AB]}{[A][B]}\,\frac{1}{V}
    \qquad \forall\,\tilde{\mathbf{S}}.
\end{equation}
Since, by definition, the configurations $\mathbf{S} \neq \tilde{\mathbf{S}}$ contribute negligibly~\footnote{Note that, in Eq.~\eqref{eq:rho_S}, the solvent-configuration-dependent single-molecule partition functions $z_{\alpha}({\bf S})$ are the terms that are assumed to be $\approx 0$ for ${\bf S}\neq\tilde{\bf S}$.}, we can integrate over the whole solvent
phase space~\footnote{%
In fact, $\frac{f(x)}{g(x)h(x)} = c \quad \forall$ $x$ $\implies$ 
$f(x)=c\,g(x)\,h(x)$ $\implies$ $\int\,f(x)\,{\rm d}x=c\int\,g(x)\,h(x)\,{\rm d}x$ $\implies$ $\frac{\int\,f(x)\,{\rm d}x}{\int\,\,g(x)\,h(x)\,{\rm d}x}=c$.}
\begin{equation}
\label{eq:K_ratio}
    \frac{\displaystyle\int\mathrm{d}\mathbf{S}\;
          z_{AB}(\mathbf{S})}
         {\displaystyle\int\mathrm{d}\mathbf{S}\;
          z_A(\mathbf{S})\,z_B(\mathbf{S})}
    = \frac{[AB]}{[A][B]}\,\frac{1}{V}.
\end{equation}
By substituting the definitions of  Eq.~\eqref{eq:z_solvent}, we obtain
\begin{equation}
\label{eq:K_explicit}
    \frac{\displaystyle
          \int\mathrm{d}\mathbf{S}
          \int_b\mathrm{d}\mathbf{a}\,\mathrm{d}\mathbf{b}\;
          e^{-\beta h_{ABS}}}
         {\displaystyle
          \int\mathrm{d}\mathbf{S}
          \int_u\mathrm{d}\mathbf{a}\,\mathrm{d}\mathbf{b}\;
          e^{-\beta h_{ABS}}}
    = \frac{[AB]}{[A][B]}\,\frac{1}{V}.
\end{equation}
Strictly speaking, the Hamiltonian in the unbound integral in the denominator is $h_{AS}+h_{BS}=h_{ABS}-h_{AB}$. However, consistent with the approximations made throughout the derivation, $h_{AB}$ is negligible for all unbound configurations, and it can be included without introducing a significant error. At finite volume, neglecting $h_{AB}$ in unbound configurations close to the bound state introduces an error of order $B^{(u)}/V$, where $B^{(u)}$ is the contribution of non-binding interactions to the second virial coefficient~\cite{JostLopez2020}. This vanishes in the thermodynamic limit. The estimators of Section~\ref{sec:restraints} of the main text, which take the unbound weight only where the partners do not interact, are free of it.

Equation~\eqref{eq:K_explicit} establishes a direct bridge between theory and explicit-solvent molecular simulations. Following the notation of the main text, it is equivalent to
\begin{equation}
\label{cs}
K=\frac{P_b}{P_u/V}.
\end{equation}

\subsection{Estimators in common geometries}
\label{sec:SI_estimators}
In this section we specialize the general expressions of the main text to the three geometries of Fig.~\ref{fig:geometries} of the main text: the extended-region estimator, Eq.~\eqref{eq:dF0R}, the hybrid estimator, Eqs.~\eqref{eq:Khybrid}--\eqref{eq:dF0hybrid}, and the single-bin estimator, Eq.~\eqref{eq:singlebin}. For each geometry we first report the continuous expressions and then derive from them the discrete forms used with simulation histograms, which specialize Eqs.~\eqref{eq:extended}--\eqref{eq:pointsgen} of the main text. We follow the notation of the main text: $\rho_R(x)$ is the probability density of the reaction coordinate $x$ in the restrained simulation, $F_R(x)=-k_{\rm B}T\log\left[\rho_R(x)\,{\rm d}x\right]$ the corresponding free energy profile, Eq.~\eqref{eq:Fslab}, $S_R(x)$ the measure of the section of the restrained region at fixed $x$, Eq.~\eqref{eq:rhoRunb}, and $V_{R,u}$ the volume of the unbound state in the restrained simulation, Eq.~\eqref{eq:VR}. Throughout, the restraint is assumed to leave the bound state untouched. Otherwise, the term $\Delta F_{\rm corr}$ of Eq.~\eqref{eq:dF0Rcorr} of the main text is added to each expression for $\Delta F_0$ below. The single-bin expressions are reported for comparison with the literature and with our results. As discussed in Section~\ref{sec:pmf} of the main text, the single-bin estimator is not a valid estimator of the binding constant, since it replaces the weight of the bound state by that of a single bin, so that its result depends on the bin width.

In the homogeneous unbound region, Eq.~\eqref{eq:rhoRunb} of the main text gives
\begin{equation}
\label{SIrhounb}
    \rho_R(x)=\bar\rho_R\,S_R(x),
    \qquad
    \bar\rho_R=\frac{\mathcal{N}}{\mathcal{N}_R}\,\bar\rho,
\end{equation}
where $\bar\rho_R$ is the constant density of the relative position per unit volume in the unbound region of the restrained system. In the spherical and circular cases, $S_R(x)$ grows with $x$, and it is convenient to factor it out of $\rho_R(x)$ by introducing a radial distribution function that approaches one in the unbound region. In the cylindrical case with a constant cross-section, $S_R$ is constant and $\rho_R(x)$ itself approaches a plateau.

\subsubsection{Spherical symmetry}
\label{SIsph}
\paragraph{Continuous expressions.}
Here $\boldsymbol{\xi}=(r,\theta,\varphi)$, $x=r$ and $|J|=r^2\sin\theta$. The restraint confines $r$ within a sphere of radius $R$ and acts only within a thin shell at its boundary, so that, for any $r$ inside the sphere and outside this shell,
\begin{equation}
\label{SIsphS}
    S_R(r)=\int_0^{\pi}{\rm d}\theta\int_0^{2\pi}{\rm d}\varphi\;r^2\sin\theta=4\pi r^2.
\end{equation}
Taking the unbound state as the shell $r\in[r_u^*,r_u^*+\ell)$ inside the sphere, Eq.~\eqref{eq:VR} of the main text gives
\begin{equation}
\label{SIsphV}
    V_{R,u}=\int_{r_u^*}^{r_u^*+\ell}{\rm d}r\,4\pi r^2=\frac{4}{3}\pi\left[\left(r_u^*+\ell\right)^3-\left(r_u^*\right)^3\right].
\end{equation}
We define the radial distribution function in the restrained simulation as
\begin{equation}
\label{SIg3cont}
    g_{3,R}(r)=\frac{\rho_R(r)}{4\pi r^2\,C_v},
\end{equation}
with $C_v=\bar\rho_R$, so that, by Eq.~\eqref{SIrhounb}, $g_{3,R}(r)=1$ in the unbound region. $C_v$ is the volumetric concentration of the free partner in the restrained system. Expressing the statistical weights through Eq.~\eqref{SIg3cont}, the extended-region estimator, Eq.~\eqref{eq:dF0R} of the main text, reads
\begin{equation}
\label{SIsphext}
    \Delta F_0=-k_{\rm B}T\log\frac{\int{\rm d}r\,I_b(r)\,r^2\,g_{3,R}(r)}{\int{\rm d}r\,I_u(r)\,r^2\,g_{3,R}(r)}-k_{\rm B}T\log\left(V_RC_0\right).
\end{equation}
With $S_R(r_u^*)=4\pi r_u^{*2}$ and $\rho_R(r_u^*)=4\pi r_u^{*2}C_v\,g_{3,R}(r_u^*)$, the hybrid estimator, Eq.~\eqref{eq:Khybrid} of the main text, becomes
\begin{equation}
\label{SIsphhyb}
    K=\frac{P_{R,b}}{C_v\,g_{3,R}(r_u^*)}=\int{\rm d}r\,I_b(r)\,4\pi r^2\,\frac{g_{3,R}(r)}{g_{3,R}(r_u^*)},
\end{equation}
where the second equality holds when the bound state is defined in terms of $r$. Since $g_{3,R}(r_u^*)=1$, Eq.~\eqref{SIsphhyb} is the integral of the radial distribution function over the bound state~\cite{Jorgensen1989,deJong2011,DuboueDijon2021}, Eq.~\eqref{eq:Kg3} of the main text. Eq.~\eqref{SIsphext} reduces to the same result, since $\int{\rm d}r\,I_u(r)\,4\pi r^2\,g_{3,R}(r)=V_{R,u}$ when $g_{3,R}=1$ over the unbound shell. Finally, the single-bin estimator, Eq.~\eqref{eq:singlebin} of the main text, reads
\begin{equation}
\label{SIsphsb}
    \Delta F_0=-k_{\rm B}T\log\frac{r_b^{*2}\,g_{3,R}(r_b^*)}{r_u^{*2}\,g_{3,R}(r_u^*)}-k_{\rm B}T\log\left(4\pi r_u^{*2}\,{\rm d}r\,C_0\right)
    =-k_{\rm B}T\log\left(4\pi r_b^{*2}\,\frac{g_{3,R}(r_b^*)}{g_{3,R}(r_u^*)}\,{\rm d}r\,C_0\right).
\end{equation}
The reference point $r_u^*$, and the unbound shell in Eq.~\eqref{SIsphV}, must lie inside the sphere, outside the shell in which the restraint acts, so that Eq.~\eqref{SIsphS} holds there.

\paragraph{Discrete expressions.}
On a mesh of bin width $\Delta r$, the radial distribution function is estimated as
\begin{equation}
\label{eq.g3SI}
    g_3(r_k)=\frac{\text{\# samples } r\in[r_k,r_k+\Delta r)}{4\pi r_k^2\,C_v\,M\,\Delta r},
\end{equation}
where $M$ is the total number of samples, so that the statistical weight of the $k$th bin is
\begin{equation}
\label{SIsphPk}
    P_{R,k}=4\pi C_v\,r_k^2\,g_3(r_k)\,\Delta r,
\end{equation}
and the free energy profile, Eq.~\eqref{eq:Afmd} of the main text, is
\begin{equation}
\label{eq.3dg3SI}
    F_R(r_k)=-k_{\rm B}T\log g_3(r_k)-k_{\rm B}T\log\left(4\pi C_v\,r^2_k\,\Delta r\right).
\end{equation}
Eq.~\eqref{SIsphPk} approximates the volume of the $k$th spherical shell by $4\pi r_k^2\Delta r$, which is accurate for $r_k\gg\Delta r$. The exact shell volume, $\frac{4}{3}\pi[(r_k+\Delta r)^3-r_k^3]$, can be used instead. Substituting Eq.~\eqref{SIsphPk} into Eqs.~\eqref{eq:extended}--\eqref{eq:pointsgen} of the main text, or, equivalently, discretizing Eqs.~\eqref{SIsphext}--\eqref{SIsphsb}, gives the extended-region estimator
\begin{equation}
\label{dF0sph}
    \Delta F_0=-k_{\rm B}T\log\frac{\sum_{k\in b}g_3(r_k)\,r_k^2}{\sum_{k\in u}g_3(r_k)\,r_k^2}-k_{\rm B}T\log\left(V_{R,u} C_0\right),
\end{equation}
with $V_{R,u}$ from Eq.~\eqref{SIsphV}, the hybrid estimator
\begin{equation}
\label{dF0sphhyb}
    \Delta F_0=-k_{\rm B}T\log\frac{\sum_{k\in b}g_3(r_k)\,r_k^2}{g_3(r_{k^*})\,r_{k^*}^2}-k_{\rm B}T\log\left(4\pi r_{k^*}^2\,\Delta r\,C_0\right),
\end{equation}
where $k^*$ is the bin centered at $r_u^*$, and the single-bin estimator
\begin{equation}
\label{dF0sphsb}
    \Delta F_0=-k_{\rm B}T\log\frac{g_3(r_{b}^*)\,r_{b}^{*2}}{g_3(r_{k^*})\,r_{k^*}^2}-k_{\rm B}T\log\left(4\pi r_{k^*}^2\,\Delta r\,C_0\right).
\end{equation}
The concentration $C_v$ cancels in all three expressions. Any nominal value can therefore be used in Eq.~\eqref{eq.g3SI}.

\subsubsection{Circular symmetry}
\label{SIcirc}
\paragraph{Continuous expressions.}
For two molecules diffusing within a membrane, the relative position is two-dimensional, $\boldsymbol{\xi}=(r,\varphi)$ in the plane of the membrane, with $x=r$ and $|J|=r$. The derivation of the main text carries over with volumes replaced by areas and the standard concentration $C_0$ by a standard surface concentration $C^0_s$, so that $K$ has the dimensions of an area and $\Delta F_0=-k_{\rm B}T\log(KC^0_s)$. For a restraint confining $r$ within a circle of radius $R$,
\begin{equation}
\label{SIcircS}
    S_R(r)=\int_0^{2\pi}{\rm d}\varphi\;r=2\pi r,
\end{equation}
and, for the unbound state taken as the ring $r\in[r_u^*,r_u^*+\ell)$,
\begin{equation}
\label{ARu}
    A_{R,u}=\pi\left[\left(r_u^*+\ell\right)^2-\left(r_u^*\right)^2\right].
\end{equation}
With the two-dimensional radial distribution function
\begin{equation}
\label{SIg2cont}
    g_{2,R}(r)=\frac{\rho_R(r)}{2\pi r\,C_s},
\end{equation}
where the surface concentration $C_s$ is such that $g_{2,R}(r)=1$ in the unbound region, the extended-region, hybrid and single-bin estimators read
\begin{equation}
\label{SIcircext}
    \Delta F_0=-k_{\rm B}T\log\frac{\int{\rm d}r\,I_b(r)\,r\,g_{2,R}(r)}{\int{\rm d}r\,I_u(r)\,r\,g_{2,R}(r)}-k_{\rm B}T\log\left(A_RC^0_s\right),
\end{equation}
\begin{equation}
\label{SIcirchyb}
    K=\frac{P_{R,b}}{C_s\,g_{2,R}(r_u^*)}=\int{\rm d}r\,I_b(r)\,2\pi r\,\frac{g_{2,R}(r)}{g_{2,R}(r_u^*)},
\end{equation}
\begin{equation}
\label{SIcircsb}
    \Delta F_0=-k_{\rm B}T\log\left(2\pi r_b^*\,\frac{g_{2,R}(r_b^*)}{g_{2,R}(r_u^*)}\,{\rm d}r\,C^0_s\right).
\end{equation}

\paragraph{Discrete expressions.}
On a mesh of bin width $\Delta r$,
\begin{equation}
\label{eq.g2}
    g_2(r_k)=\frac{\text{\# samples } r\in[r_k,r_k+\Delta r)}{2\pi r_k\,C_s\,M\,\Delta r},
    \qquad
    P_{R,k}=2\pi C_s\,r_k\,g_2(r_k)\,\Delta r,
\end{equation}
and the three estimators become
\begin{equation}
\label{eq.dA0_2D_rest}
    \Delta F_0=-k_{\rm B}T\log\frac{\sum_{k\in b}g_2(r_k)\,r_k}{\sum_{k\in u}g_2(r_k)\,r_k}-k_{\rm B}T\log\left(A_{R,u}\,C^0_s\right),
\end{equation}
\begin{equation}
\label{dF0circhyb}
    \Delta F_0=-k_{\rm B}T\log\frac{\sum_{k\in b}g_2(r_k)\,r_k}{g_2(r_{k^*})\,r_{k^*}}-k_{\rm B}T\log\left(2\pi r_{k^*}\,\Delta r\,C^0_s\right),
\end{equation}
\begin{equation}
\label{dF0circsb}
    \Delta F_0=-k_{\rm B}T\log\frac{g_2(r_{b}^*)\,r_{b}^*}{g_2(r_{k^*})\,r_{k^*}}-k_{\rm B}T\log\left(2\pi r_{k^*}\,\Delta r\,C^0_s\right).
\end{equation}
As in the spherical case, $C_s$ cancels and any nominal value can be used.

\subsubsection{Cylindrical symmetry}
\label{SIcyl}
\paragraph{Continuous expressions.}
Here $\boldsymbol{\xi}=(z,s,\varphi)$, with $z$ the displacement along the axis directed from the binding site to the bulk, $s$ the lateral distance from the axis, $x=z$ and $|J|=s$. The restraint confines the lateral displacement within a cross-section of area $\Sigma(z)$ in the unbound region, so that
\begin{equation}
\label{SIcylS}
    S_R(z)=\int_0^{2\pi}{\rm d}\varphi\int{\rm d}s\;s\,I_R(z,s,\varphi)=\Sigma(z),
\end{equation}
which equals $\pi R^2$ for a cylinder of radius $R$. For the unbound state taken as $z\in[z_u^*,z_u^*+\ell)$,
\begin{equation}
\label{VRucyl}
    V_{R,u}=\int_{z_u^*}^{z_u^*+\ell}{\rm d}z\,\Sigma(z),
\end{equation}
which reduces to $V_{R,u}=\Sigma\ell$ for a constant cross-section. In this case, by Eq.~\eqref{SIrhounb}, $\rho_R(z)=\bar\rho_R\,\Sigma$ is itself constant in the unbound region and provides the convergence diagnostic. For restraints whose cross-section varies along $z$, such as funnel-shaped ones, the unbound state and the reference point $z_u^*$ are most conveniently placed in the region of constant cross-section. With a constant cross-section, the extended-region estimator reads
\begin{equation}
\label{SIcylext}
    \Delta F_0=-k_{\rm B}T\log\frac{\int{\rm d}z\,I_b(z)\,\rho_R(z)}{\int{\rm d}z\,I_u(z)\,\rho_R(z)}-k_{\rm B}T\log\left(\Sigma\ell C_0\right),
\end{equation}
the hybrid estimator
\begin{equation}
\label{Kcylhybrid}
    K=\Sigma\,\frac{P_{R,b}}{\rho_R(z_u^*)}=\Sigma\int{\rm d}z\,I_b(z)\,e^{-\beta\left[F_R(z)-F_R(z_u^*)\right]},
\end{equation}
which, with $\Sigma=\pi R_{\rm cyl}^2$, is the expression of Refs.~\cite{Allen2004,Woo2005,Limongelli2013}, and the single-bin estimator
\begin{equation}
\label{SIcylsb}
    \Delta F_0=F_R(z_b^*)-F_R(z_u^*)-k_{\rm B}T\log\left(\Sigma\,{\rm d}z\,C_0\right).
\end{equation}

\paragraph{Discrete expressions.}
On a mesh of bin width $\Delta z$, with $P_{R,k}$ the normalized histogram of $z$ and $F_R(z_k)=-k_{\rm B}T\log P_{R,k}$, Eqs.~\eqref{SIcylext}--\eqref{SIcylsb} become
\begin{equation}
\label{dF0cylext}
    \Delta F_0=-k_{\rm B}T\log\frac{\sum_{k\in b}P_{R,k}}{\sum_{k\in u}P_{R,k}}-k_{\rm B}T\log\left(\Sigma\ell C_0\right),
\end{equation}
\begin{equation}
\label{dF0cylhyb}
    \Delta F_0=-k_{\rm B}T\log\frac{\sum_{k\in b}P_{R,k}}{P_{R,k^*}}-k_{\rm B}T\log\left(\Sigma\,\Delta z\,C_0\right),
\end{equation}
\begin{equation}
\label{dF0cylsb}
    \Delta F_0=F_R(z_b^*)-F_R(z_{k^*})-k_{\rm B}T\log\left(\Sigma\,\Delta z\,C_0\right),
\end{equation}
which are Eqs.~\eqref{eq.K_1D_rest}--\eqref{eq:points} of the main text.

\section{Applications}
\subsection{Lennard-Jones dimer}
\label{subsec:LJ}
We performed molecular dynamics (MD) simulations of Lennard-Jones (LJ) particles in the canonical ($NVT$) ensemble using LAMMPS~\cite{lammps}. Particles interact through the short-ranged pairwise LJ potential
\begin{equation}
    V_{\alpha\beta}(r)=4\varepsilon_{\alpha\beta}
    \left[\left(\frac{\sigma}{r}\right)^{12}
    -\left(\frac{\sigma}{r}\right)^6\right]\theta(2.5\,\sigma-r),
\end{equation}
where $\alpha,\beta \in \{P,S\}$ label the particle types, $r$ is the inter-particle distance and $\theta(2.5\,\sigma-r)$ is the Heaviside step function, setting a cutoff for the interaction at $2.5\,\sigma$. We set  $\sigma=1$. We also set the mass of both types of particles to $m=1$. Two type-P
particles (the dimer, $\varepsilon_{PP} = 5$) were immersed in a bath of $39{,}998$ type-S solvent particles ($\varepsilon_{SS} = 1$), with cross-interaction parameter $\varepsilon_{SP} = 1$. The system was simulated at
number density $\rho_0 = 0.8~\sigma^{-3}$ (cubic simulation box of side $L = (N/\rho_0)^{1/3} \approx 36.8\,\sigma$) and temperature $T = 0.9\,\varepsilon_{SS}/k_{\rm B}$, corresponding to a dense liquid phase of the solvent. One type-P particle was held fixed at the centre of the box. The second was confined by the potential
\begin{equation}
    U_{\rm wall}(r) = \frac{1}{2}k(r - R)^2\,\theta(R - r)\,\theta(r-R+0.5\,\sigma),
\end{equation}
with elastic constant $k = 50\,\varepsilon_{SS}/\sigma^2$, where $r$ is the dimer inter-particle distance and $\theta(x)$ is the Heaviside step function. The wall pushes the particle inward only in a 0.5~$\sigma$-wide shell just inside $R$. Three values of the restraint radius were considered: $R =10\,\sigma, 12\,\sigma, 14\,\sigma$. Temperature was maintained using the
stochastic velocity rescaling thermostat~\cite{Bussi2007} with time constant $\tau_T = 1\,\tau_{\rm LJ}$, where $\tau_{\rm LJ}=\sqrt{m\,\sigma^2/\varepsilon_{SS}}$. Equations of motion were integrated using the velocity-Verlet algorithm with time step $\Delta t = 0.005\,\tau_{\rm LJ}$. The simulations were run for $10^9$ steps, yielding $2\times10^6$ samples of the dimer distance $r$ per simulation.

\begin{figure}
  \centering
\includegraphics[width=\linewidth]{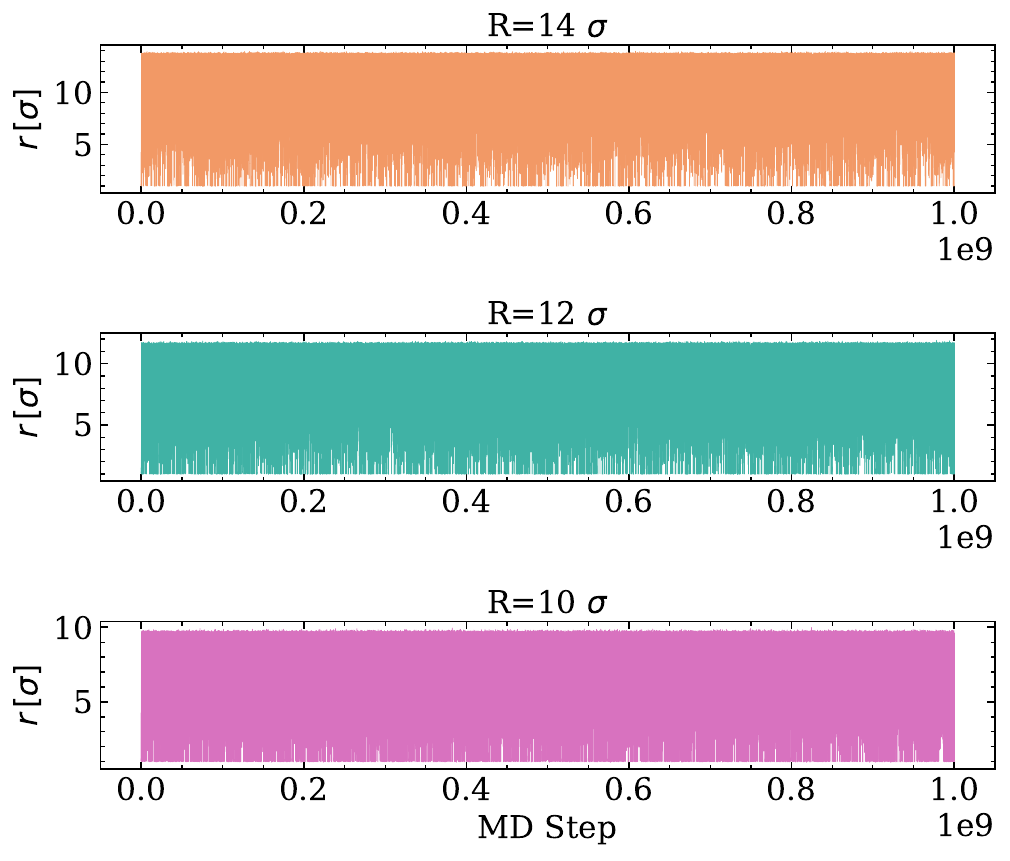}
  \caption{Dimer separation $r$ (in units of $\sigma$) as a function of MD
    step for the three LJ simulations, performed with spherical
    restraining walls of radius $R = 10\,\sigma$, $12\,\sigma$, $14\,\sigma$ (bottom to top). The dimer reversibly explores the full range
    between the bound state ($r \approx 1\,\sigma$) and the wall ($r \approx R-0.5\,\sigma$),
    confirming adequate sampling of both the bound and unbound regions.}
  \label{fig:LJ_traj}
\end{figure}

Figure~\ref{fig:LJ_traj} shows the dimer inter particle distance $r$ as a function of MD step for the three simulations.

Table~\ref{tab:lj_results} reports the binding affinity obtained using the following definitions of the bound and unbound states:
\begin{itemize}
    \item \textit{Estimate 1}. Bound region: $r\in[0,1.8\,\sigma]$. Unbound region: $r\in[4\,\sigma,R-0.5\,\sigma]$. The unbound region volume $V_{R,u}=\frac{4}{3}\pi[(R-0.5)^3-(4)^3]\,\sigma^3$ is equal to $V_{R,u}=3{,}323\,\sigma^3,\,6{,}103\,\sigma^3,\,10{,}038\,\sigma^3$, respectively for $R=10\,\sigma,\,12\,\sigma,\,14\,\sigma$.
    \item \textit{Estimate 2}. Bound region: $r\in[0,1.8\,\sigma]$. Unbound region: the single bin $r\in[5\,\sigma-\Delta r,\,5\,\sigma)$. Bin width $\Delta r=0.05\,\sigma$. The unbound region volume is $V_{R,u}=\frac{4}{3}\pi[5^3-(5-\Delta r)^3]\,\sigma^3=15.55\,\sigma^3$.
    \item \textit{Estimate 3}. Bound region: the single bin $r\in[1.1\,\sigma,\,1.1\,\sigma+\Delta r)$, at the maximum of the bound-state peak. Unbound region: as in Estimate 2. Bin width $\Delta r=0.05\,\sigma$. The unbound region volume is $V_{R,u}=15.55\,\sigma^3$.
\end{itemize}
Estimates 1 and 2 agree with each other within statistical uncertainty. Since Estimate 1 and the estimates reported in Table~\ref{tab:LJ} of the main text are computed from the same frames, their difference is systematic: Estimate 1 is $\approx 0.08\,\varepsilon_{\rm SS}$ less negative for all three radii. The difference originates entirely from the definition of the bound state: the region $1.8\,\sigma<r<2.5\,\sigma$, included in the bound state in the main text, carries $\approx 9\%$ of the bound-state weight, while the density of the unbound state agrees between the two definitions to within $0.2\%$. Estimate 3 -- which reduces the bound state to a single bin in the histogram -- deviates from Estimate 2 by $\approx 1.1\,\varepsilon_{\rm SS}$ ($\approx 1.2\,k_{\rm B}T$). Estimate 3 is the single-bin estimator, which is not a valid estimator of the binding constant (Section~\ref{sec:pmf} of the main text).
\subsection{Cucurbit[7]uril/adamantanol host–guest complex}
\label{subsec:cb7}
\begin{figure}
    \centering
    \includegraphics[width=0.8\textwidth]{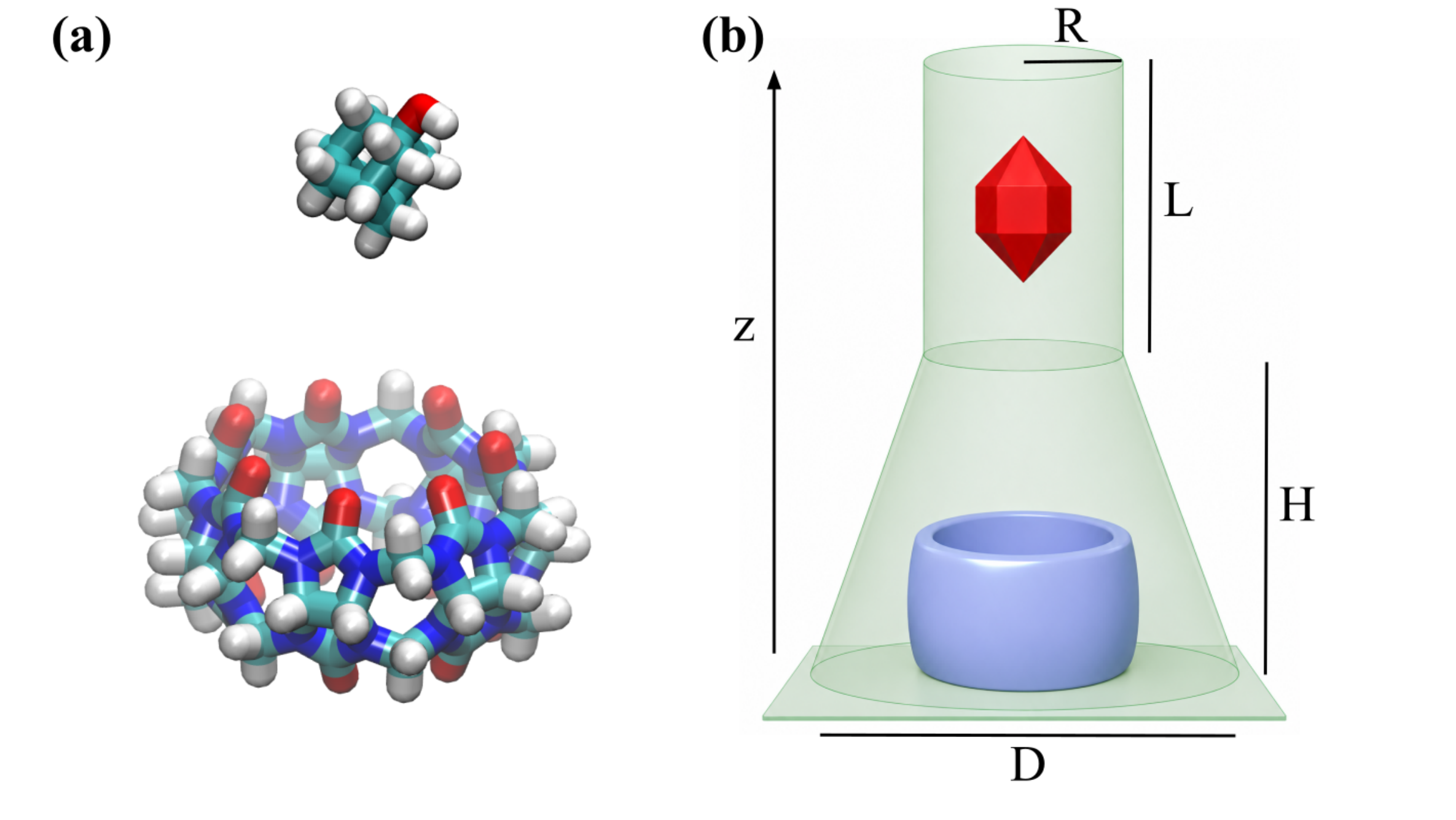}
    \caption{
        (a)~ Atomistic model of 1-adamantanol (top) and CB7 (bottom) in explicit water (omitted for clarity) used in simulations. Atoms are colored by element: carbon (cyan), oxygen (red), nitrogen (blue), and hydrogen (white).
        (b)~Schematic of the restraint adopted in simulations. A truncated cone (base diameter $D$, height $H$) encloses the CB7 binding site and joins a cylindrical section (radius $R$, height $L$) extending into the bulk solvent. An additional harmonic wall is placed at the bottom of CB7, preventing escape of the guest from that side. The reaction coordinate $z$ (vertical arrow) is the projection of the ligand-CB7 centre-of-mass displacement onto the CB7 symmetry axis.
    }
    \label{fig:cb7_setting}
\end{figure}

The system used in simulations is shown in Fig.~\ref{fig:cb7_setting}a. The setting, force field parameters, and initial equilibrated configurations were taken from Ref.~\cite{Vijay2025}. Simulations were performed in the isothermal--isobaric ($NPT$) ensemble at $T = 300$~K and $P = 1$~bar, using GROMACS 2022~\cite{Abraham2015} patched with PLUMED 2.8~\cite{Tribello2014}. The temperature was maintained using the stochastic velocity rescaling thermostat~\cite{Bussi2007} with time constant $\tau_T=0.1$~ps. The pressure was controlled using the isotropic Parrinello-Rahman barostat~\cite{Parrinello1980,Parrinello1981} with time constant $\tau_P=2.0$~ps and compressibility of $4.5\times 10^{-5}$~bar$^{-1}$. All covalent bonds involving hydrogen atoms were constrained to their equilibrium length using the LINCS algorithm~\cite{hess1997}. Short range Van der Waals and electrostatic interactions were computed within a cutoff radius of 1.2~nm. Long range electrostatic interactions were computed using the Particle Mesh Ewald scheme~\cite{darden1993pme,essmann1995spme} as implememnted in GROMACS, setting \textit{pme\_order}=4 and \textit{fourierspacing}=0.16~nm.

Well-tempered metadynamics~\cite{Barducci2008} simulations were performed while restraining the center-of-mass of the guest molecule within the volume shown in Fig.~\ref{fig:cb7_setting}b. The volume consists of a conical section, leaving the binding site unperturbed, and a cylindrical section of radius $R$ extending into the bulk solvent for a distance $L=2.5$~nm. The truncated cone has a basis of diameter $D\approx2$~nm and a height $H\approx1.2$~nm. The surface acts as a repulsive harmonic potential with elastic constant $k = 3.51\times 10^4$~kJ~mol$^{-1}$~nm$^{-2}$. An additional repulsive wall with elastic constant $k_{\rm wall} = 5\times10^5$~kJ~mol$^{-1}$~nm$^{-2}$ was placed on one side of CB7 to restrict (un)binding to a single portal, accelerating convergence without affecting the computed binding affinity. We note that the conical component of the restraint potential may alter the
statistical weight of configurations in which 1-adamantanol is only
partially desolvated near the portal, relative to an unrestrained simulation. We considered three distinct cylinder radii of size $R = 0.1,\,0.2,\,0.3\,\mathrm{nm}$. Following Ref.~\cite{Vijay2025}, two collective variables were used in the metadynamics simulation: the component $z$ of the host-guest centre-of-mass separation along the
CB7 symmetry axis, and the angle $\theta$ describing the orientation of the hydroxyl C--O bond vector relative to the CB7 symmetry axis. The latter was included to enhance sampling of the orientational degree of freedom of the ligand inside the cavity. Well-tempered metadynamics as implemented in PLUMED~\cite{Tribello2014} was run with the following parameters: Gaussian kernels deposited every PACE$=500$~steps, BIASFACTOR$=20$, initial Gaussian HEIGHT$=1.0$~kJ/mol, Gaussian standard deviations SIGMA$_z=0.1$~nm and SIGMA$_\theta=0.15$~rad for the two collective variables, respectively. The bias potential was stored on a $150\times150$ grid in $(z,\theta)$, spanning
$z\in[-0.7,2.7]$~nm and $\theta\in(-\pi,\pi]$, and interpolated between grid
points, with periodicity enforced in $\theta$. 

Equations of motion were integrated using the leap-frog algorithm with a time step of $\Delta t = 2$~fs. Each simulation was run for a total duration of $\approx700$~ns. 

\begin{figure}
  \centering
  \includegraphics[width=\linewidth]{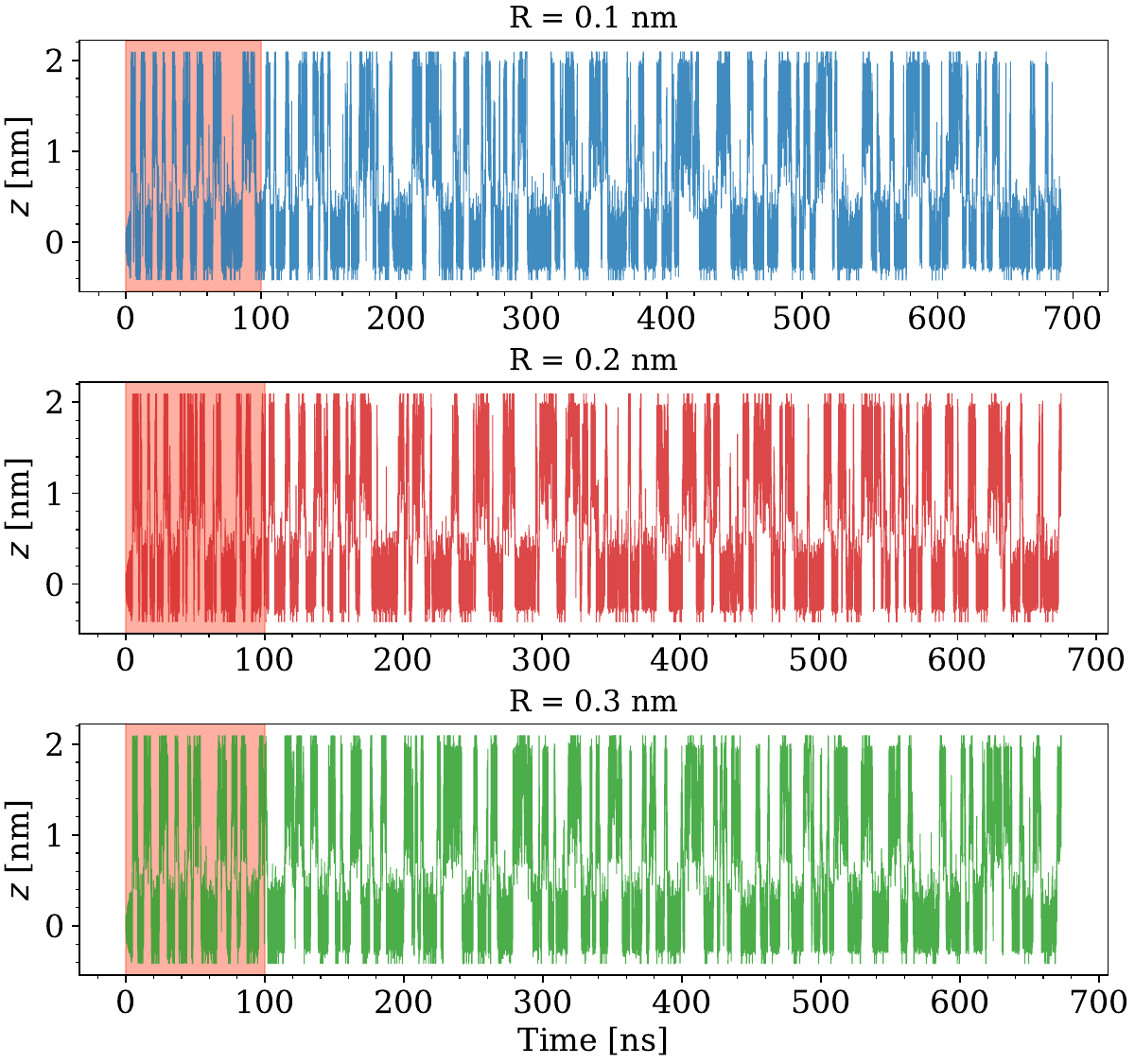}
  \caption{Trajectory of the reaction coordinate $z$ (host--guest
    separation along the cylinder axis) for the CB7--1-adamantanol system,
    for cylindrical restraints of radius $R = 0.1$, $0.2$, and $0.3$~nm
    (top to bottom). The guest reversibly samples the bound
    ($z \approx 0$~nm) and unbound ($z \gtrsim 1$~nm) regions. The shaded
    area indicates the initial transient prior to the onset of the adiabatic regime of well-tempered metadynamics, which was discarded from the analysis
    (see Fig.~\ref{fig:cb7_ct}).}
  \label{fig:cb7_z}
\end{figure}

\begin{figure}
  \centering
  \includegraphics[width=\linewidth]{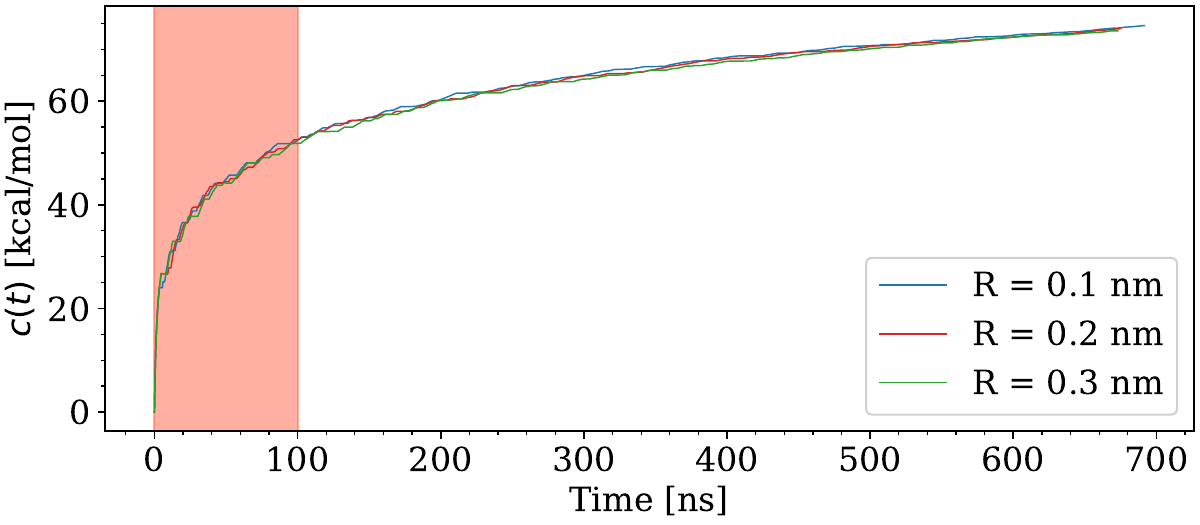}
  \caption{Time evolution of the well-tempered metadynamics factor $c(t)$~\cite{Tiwary2015a} in the three simulations performed using cylinder radii of $R = 0.1$, $0.2$, $0.3$~nm. The shaded area marks the discarded initial transient prior to the onset of the adiabatic regime.}
  \label{fig:cb7_ct}
\end{figure}

Figure~\ref{fig:cb7_z} shows the trajectory of the reaction coordinate $z$ for the three simulations. To determine the portion of each trajectory to discard as transient, we monitored the metadynamics reweighting factor $c(t)$~\cite{Tiwary2015a}, shown in Fig.~\ref{fig:cb7_ct}: the shaded region marks the initial,  discarded segment. The same interval is shaded in Fig.~\ref{fig:cb7_z} for consistency.

Table~\ref{tab:cb7_daresults} reports the values of the binding affinity obtained defining the bound state as the interval $z\in[-0.3,1.5)\,\textrm{nm}$ and the unbound region as the complementary interval $z\in[1.5,2.0]\,\textrm{nm}$. The unbound region volume is computed as $V_{R,u}=\pi R^2\ell$ with $\ell=0.5\,\textrm{nm}$, yielding $V_{R,u}=0.0157,\,0.0628,\,0.1414\,\textrm{nm}^3$, respectively for $R=0.1,\,0.2,\,0.3\,\textrm{nm}$. Note that the bound state includes regions that are affected by the restraint. However, given that the statistical weight of the bound state is dominated by the main bound pose (where the restraint is inactive), the final estimate is largely unaffected and agrees with the values obtained using the extended-region and hybrid-estimators reported in Table~\ref{tab:cb7_results} of the main text.
\subsection{Galactonate–DgoT ligand–protein complex}
\label{subsec:dgot}
The system setting, force field parameters, equilibration protocol, and metadynamics simulations setting are described in Ref.~\cite{Plate2026}, which provides full details. Briefly, well-tempered metadynamics~\cite{Barducci2008} was used to compute the free energy profile for the unbinding of deprotonated D-galactonate (GAL) from the inward-facing, gate-open conformation of the bacterial transporter DgoT, using approximately 10~$\mu$s of sampling at $T = 310.15$~K. To accelerate convergence, the center-of-mass of GAL was restricted to move within a smooth restraining volume ending with a cylindrical region of radius $R=0.1$~nm in the solvent bulk region. A single collective variable was used for metadynamics, corresponding to the $z$-projection of the center-of-mass displacement between DgoT's binding pocket and GAL onto the intracellular release axis. As discussed in Ref.~\cite{Plate2026}, the restraint may be active for configurations along the unbinding pathways that are however safely far away from the main bound pose (see Fig. 2 in Ref.~\cite{Plate2026}).

\begin{figure}[!t]
  \centering
  \includegraphics[width=\linewidth]{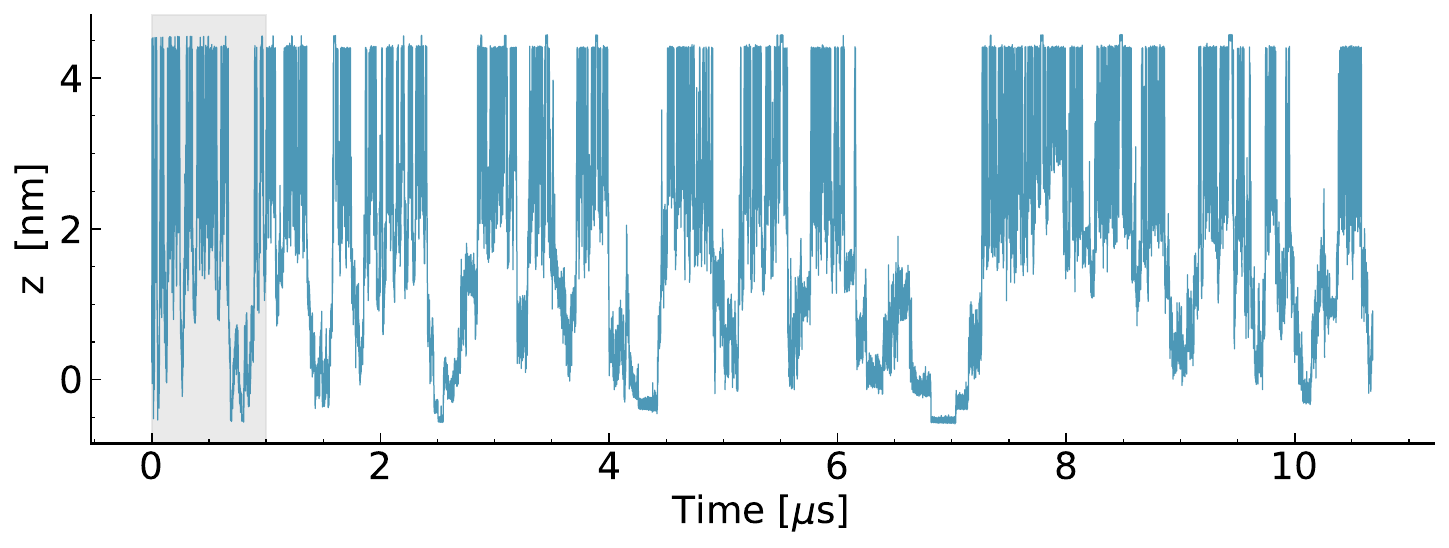}
  \caption{Trajectory of the reaction coordinate $z$ (ligand--protein separation along the protein release axis). The ligand reversibly explores the bound ($z \approx 0$~nm) and unbound
    ($z \gtrsim 2$~nm) regions. The shaded
    area indicates the initial transient prior to the onset of the adiabatic regime of well-tempered metadynamics, which was discarded from the analysis
    (see Fig.~\ref{fig:dgot_ct}).}
  \label{fig:dgot_z}
\end{figure}

\begin{figure}[!t]
  \centering
  \includegraphics[width=\linewidth]{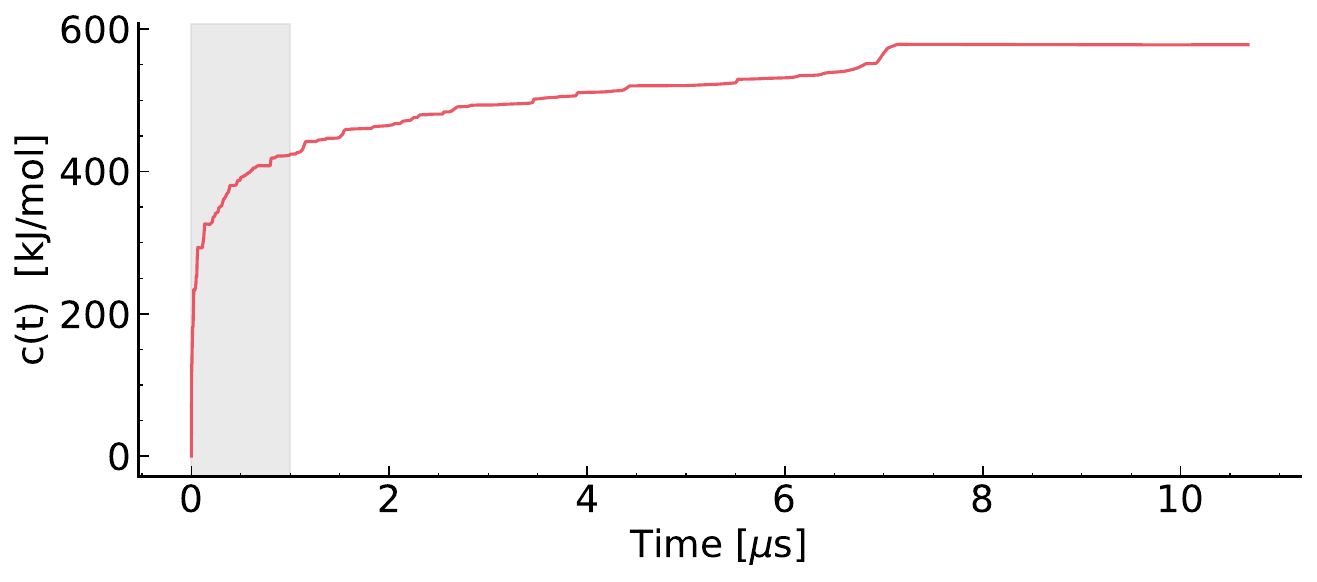}
  \caption{Trajectory of the metadynamics reweighting factor
    $c(t)$~\cite{Tiwary2015a}. The shaded area marks the discarded initial transient prior to the onset of the adiabatic regime.}
  \label{fig:dgot_ct}
\end{figure}

Figure~\ref{fig:dgot_z} shows the trajectory of the reaction coordinate $z$ that reversibly samples the bound and unbound regions over the course of the simulation. The transient discarded from the analysis was determined from the metadynamics reweightng factor $c(t)$~\cite{Tiwary2015a}, shown in Fig.~\ref{fig:dgot_ct}: the shaded interval marks the initial non-stationary regime, excluded from the analysis. The same interval is shaded in Fig.~\ref{fig:dgot_z}.

\begin{figure}[!t]
  \centering
  \includegraphics[width=\linewidth]{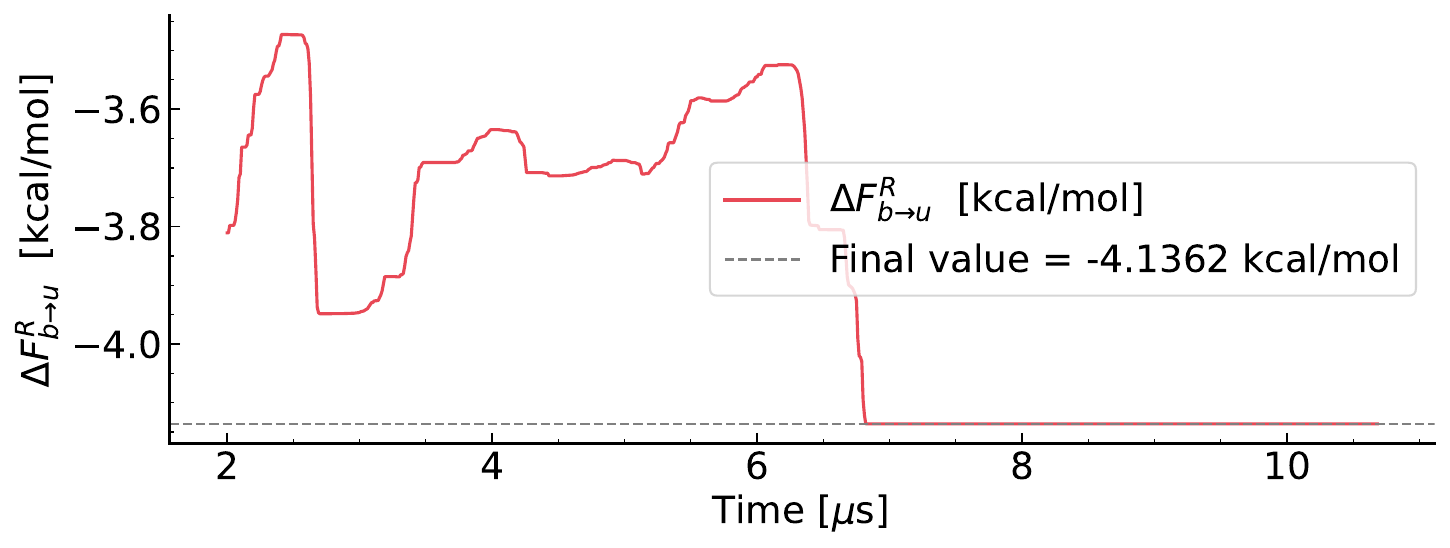}
  \caption{Running estimate of the volume-dependent binding free energy using the
    extended-region estimator (first entry of Table~\ref{tab:dgot_results} of the main text).}
  \label{fig:dgot_dF}
\end{figure}

\begin{figure}
  \centering
  \includegraphics[width=\linewidth]{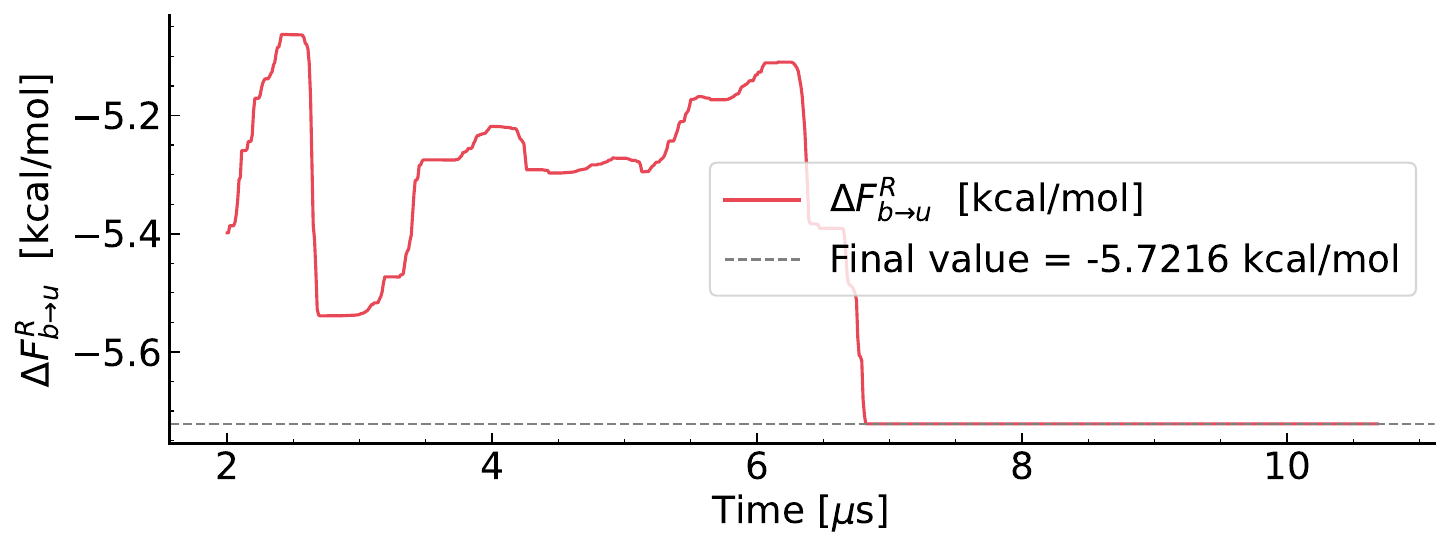}
  \caption{Running estimate of the volume-dependent binding free energy using the
    hybrid estimator (second entry of Table~\ref{tab:dgot_results} of the main text).}
  \label{fig:dgot_dF_hybrid}
\end{figure}

\begin{figure}
  \centering
  \includegraphics[width=\linewidth]{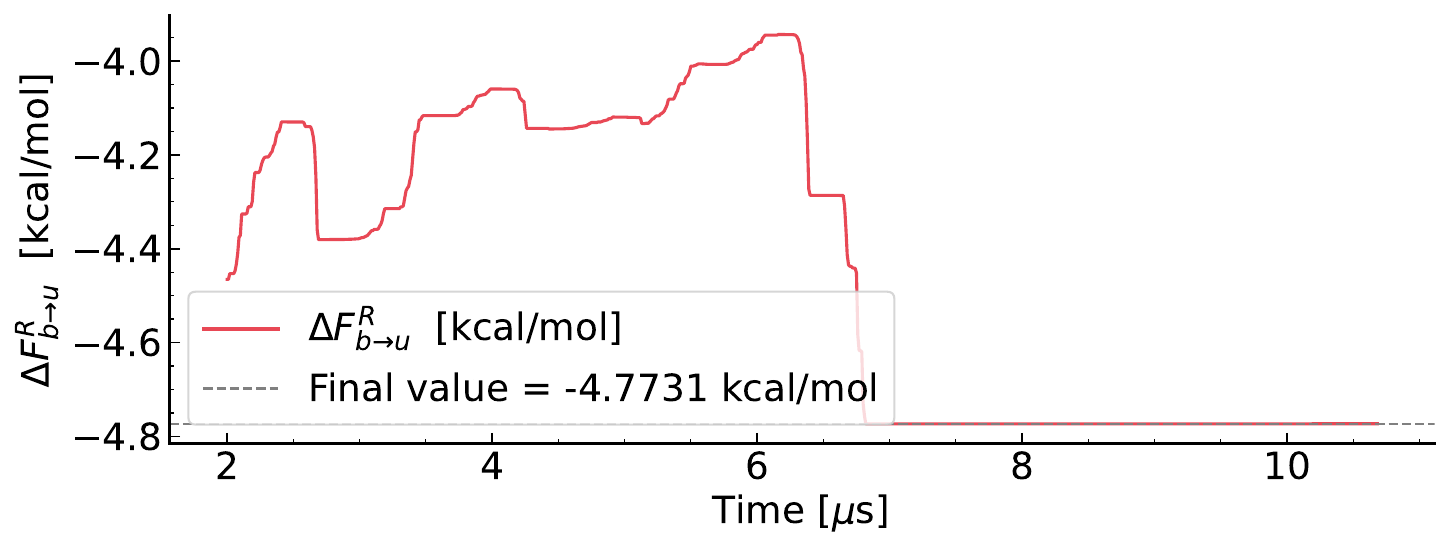}
  \caption{Running estimate of the volume-dependent binding free energy using the
    single-bin estimator (third entry of Table~\ref{tab:dgot_results} of the main text).}
  \label{fig:dgot_dF_singlebin}
\end{figure}

\begin{figure}
  \centering
  \includegraphics[width=\linewidth]{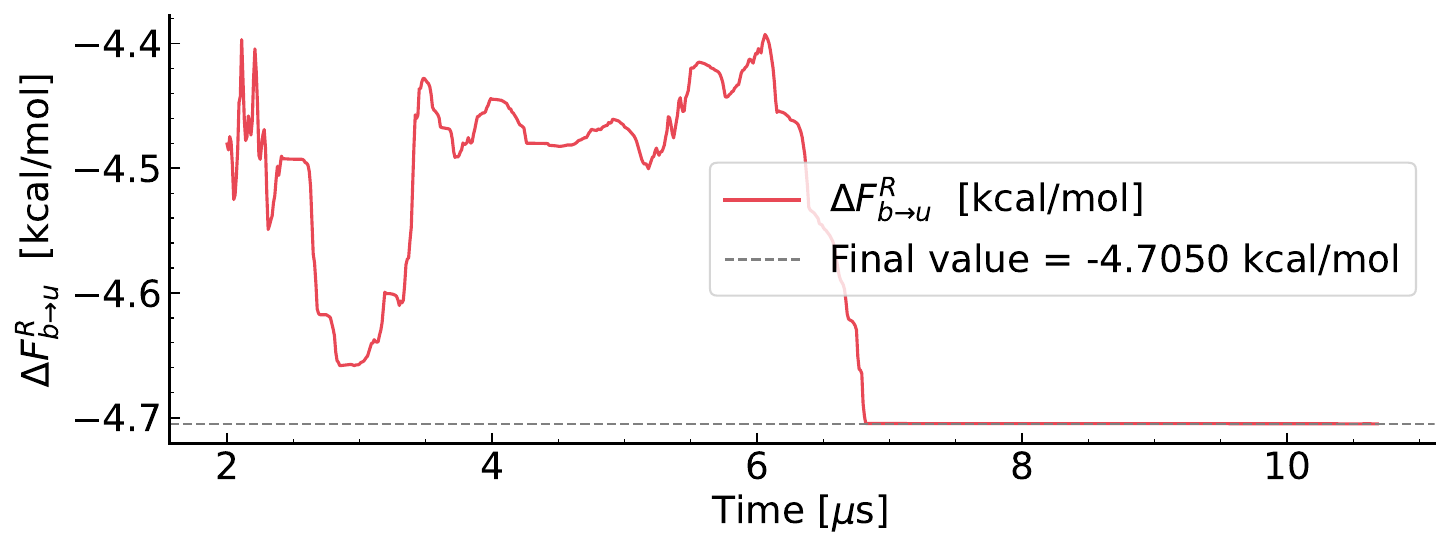}
  \caption{Running estimate of the volume-dependent binding free energy computed using the
    extended-region estimator with complementary bound and unbound states (last entry of Table~\ref{tab:dgot_results} of the main text).}
  \label{fig:dgot_dF_compl}
\end{figure}

Figures~\ref{fig:dgot_dF},~\ref{fig:dgot_dF_hybrid},~\ref{fig:dgot_dF_singlebin} show the running estimate of the binding free energy $\Delta F_{R,u\to b}$ as a function of simulation time, respectively for the extended-region, hybrid and single-bin estimators. Figure~\ref{fig:dgot_dF_compl} shows the same plot for the extended-region estimator in which the bound and unbound states are defined as complementary regions partitioning the phase space. This analysis was used to estimate the uncertainties reported in Table~\ref{tab:dgot_results} of the main text.

%
\clearpage
\clearpage
\begin{table}
\centering
\caption{
    \textbf{Binding affinity of the LJ dimer.} 
    Energies in units of $\varepsilon_{\rm SS}$. $C_0 = 1$ $\sigma^{-3}$.  Uncertainties on the last digits are $\pm 1$ SD from seven independent estimates. See text for definitions of the three different estimates.\\
}
\label{tab:lj_results}
\begin{tabular}{cccc}
\toprule
$R$ [$\sigma$] & $\Delta F_{R,u\to b}$ & $-k_{\rm B}T\log(V_{R,u} C_0)$ & $\Delta F_0$ \\
\midrule
\multicolumn{4}{l}{\textit{Estimate 1}} \\[3pt]
10 & 1.65(5) & -7.30 & -5.65(5) \\
12 & 2.20(13) & -7.84 & -5.65(13) \\
14 & 2.59(9) & -8.29 & -5.71(9) \\
\midrule
\multicolumn{4}{l}{\textit{Estimate 2}} \\[3pt]
10 & -3.17(5) & -2.47 & -5.64(5) \\
12 & -3.14(12) & -2.47 & -5.61(12) \\
14 & -3.19(10) & -2.47 & -5.66(10) \\
\midrule
\multicolumn{4}{l}{\textit{Estimate 3}} \\[3pt]
10 & -2.08(6) & -2.47 & -4.55(6) \\
12 & -2.04(12) & -2.47 & -4.51(12) \\
14 & -2.10(10) & -2.47 & -4.56(10) \\
\bottomrule
\end{tabular}
\end{table}
\clearpage
\clearpage
\clearpage
\clearpage
\begin{table}
\centering
\caption{
    \textbf{Binding affinity of the CB7/1-adamantanol complex.} 
    Energies in kcal/mol. $C_0 = 1$ mol/L.  Uncertainties on the last digits are $\pm 1$ SD from six independent estimates.\\
}
\label{tab:cb7_daresults}
\begin{tabular}{cccc}
\toprule
        $R\,[\textrm{nm}]$ & 
        $\Delta F_{R,u\to b}$ &
         $-k_{\rm B}T\log(V_{R,u} C_0)$ &
        $\Delta F_0$ \\
        \midrule
        0.1  & -20.49(28) & 2.78 & -17.71(28) \\
        0.2  & -19.70(10) & 1.95 & -17.74(10) \\
        0.3  & -19.04(20) & 1.47 & -17.57(20) \\
        \bottomrule
\end{tabular}
\end{table}
\clearpage
\clearpage
\clearpage
\clearpage
\clearpage
\clearpage
\clearpage
%
\FloatBarrier

\end{document}